\documentclass[usenatbib]{mnras}

\usepackage[T1]{fontenc}
\usepackage{ae,aecompl}

\usepackage{graphicx}	
\usepackage{amsmath}	
\usepackage{amssymb}	
\usepackage[normalem]{ulem} 
\usepackage{multirow} 
\usepackage{xcolor} 
\usepackage{lipsum} 
\usepackage{hyperref} 
\usepackage{mathtools} 
\usepackage{subcaption}
\usepackage{array} 
\usepackage{txfonts}
\usepackage{physics}
\usepackage[shortcuts]{extdash}

\newcommand{\D}{\mathrm{d}}

\DeclareRobustCommand{\VAN}[3]{#2}
\let\VANthebibliography\thebibliography
\def\thebibliography{\DeclareRobustCommand{\VAN}[3]{##3}\VANthebibliography}

\makeatletter
\def\footnoterule{\kern-3\p@
  \hrule \@width 2in \kern 2.6\p@} 
\makeatother

\title[On the Cosmic Web Elongation in FDM Cosmologies]{On the Cosmic Web Elongation in Fuzzy Dark Matter Cosmologies: Effects on Density Profiles, Shapes and Alignments of Halos}

\author[T. Dome et al.]{Tibor Dome$^{1,2}$\thanks{E-mail: td448@cam.ac.uk}, Anastasia Fialkov$^{1,2}$, Philip Mocz$^{3}$, Bj\"orn Malte Sch\"afer$^{4}$,
\newauthor
Michael Boylan-Kolchin$^{5}$, Mark Vogelsberger$^{6}$\\
$^{1}$Institute of Astronomy, University of Cambridge, Madingley Road, Cambridge, CB3 0HA, UK\\
$^{2}$Kavli Institute for Cosmology, Madingley Road, Cambridge, CB3 0HA, UK\\
$^{3}$Department of Astrophysical Sciences, Princeton University, 4 Ivy Lane, Princeton, NJ, 08544, USA\\
$^{4}$ Zentrum f\"ur Astronomie der Universit\"at Heidelberg, Astronomisches Rechen-Institut, Philosophenweg 12, 69120 Heidelberg, Germany\\
$^{5}$ Department of Astronomy, The University of Texas at Austin, 2515 Speedway, Stop C1400, Austin, TX 78712-1205, USA\\
$^{6}$ Department of Physics, Kavli Institute for Astrophysics and Space Research, Massachusetts Institute of Technology, Cambridge, MA 02139, USA
}

\date{Accepted XXX. Received YYY; in original form ZZZ}

\pubyear{2022}

\begin{document}
\label{firstpage}
\pagerange{\pageref{firstpage}--\pageref{lastpage}}
\maketitle

\begin{abstract}
The fuzzy dark matter (FDM) scenario has received increased attention in recent years due to the small-scale challenges of the vanilla Lambda cold dark matter ($\Lambda$CDM) cosmological model and the lack of any experimental evidence for any candidate particle. In this study, we use cosmological $N$-body simulations to investigate high-redshift dark matter halos and their responsiveness to an FDM-like power spectrum cutoff on small scales in the primordial density perturbations. We study halo density profiles, shapes and alignments in FDM-like cosmologies (the latter two for the first time) by providing fits and quantifying departures from $\Lambda$CDM as a function of the particle mass $m$. Compared to $\Lambda$CDM, the concentrations of FDM-like halos are lower, peaking at an $m$-dependent halo mass and thus breaking the approximate universality of density profiles in $\Lambda$CDM. The intermediate-to-major and minor-to-major shape parameter profiles are monotonically increasing with ellipsoidal radius in $N$-body simulations of $\Lambda$CDM. In FDM-like cosmologies, the monotonicity is broken, halos are more elongated around the virial radius than their $\Lambda$CDM counterparts and less elongated closer to the center. Finally, intrinsic alignment correlations, stemming from the deformation of initially spherically collapsing halos in an ambient gravitational tidal field, become stronger with decreasing $m$. At $z\sim 4$, we find a $6.4 \sigma$-significance in the fractional differences between the isotropised linear alignment magnitudes $D_{\text{iso}}$ in the $m=10^{-22}$ eV model and $\Lambda$CDM. Such FDM-like imprints on the internal properties of virialised halos are expected to be strikingly visible in the high-$z$ Universe.

\end{abstract}




\section{Introduction}
\subsection{Wave Dark Matter}
The six-parameter standard Lambda cold dark matter ($\Lambda$CDM) cosmological scenario that emerged in the late 1990s can boast about a wide range of observational successes, from the Cosmic Microwave Background \citep[CMB]{Planck_2015}, the Lyman-$\alpha$ forest \citep{Irsic_2016} and galaxy clustering \citep{Nuza_2013} to weak gravitational lensing \citep{Murata_2018}.\par

While this "minimal" approach might be justified by the good fit to the CMB data and by model selection criteria like Occam's razor, some of the theoretical arguments underpinning the $\Lambda$CDM approach lack sufficient justification \citep{DiValentino_2015}: The total neutrino mass $\sum m_{\nu} = 0.06$ eV/$c^2$ is set to the minimal value allowed by solar and terrestrial neutrino oscillation experiments while cosmological data sets are sensitive to $\sim 100$ meV variations and the dark energy equation of state is fixed to $w=-1$ so as to represent a cosmological constant rather than allow for time- and position-dependent components. Further, the effective number of light neutrino species is set to the Standard Model value of $N_{\text{eff}} \sim 3.046$ which e.g. inflationary reheating or non-standard decoupling could alter.\par

\begin{figure*}
\includegraphics[scale=0.5]{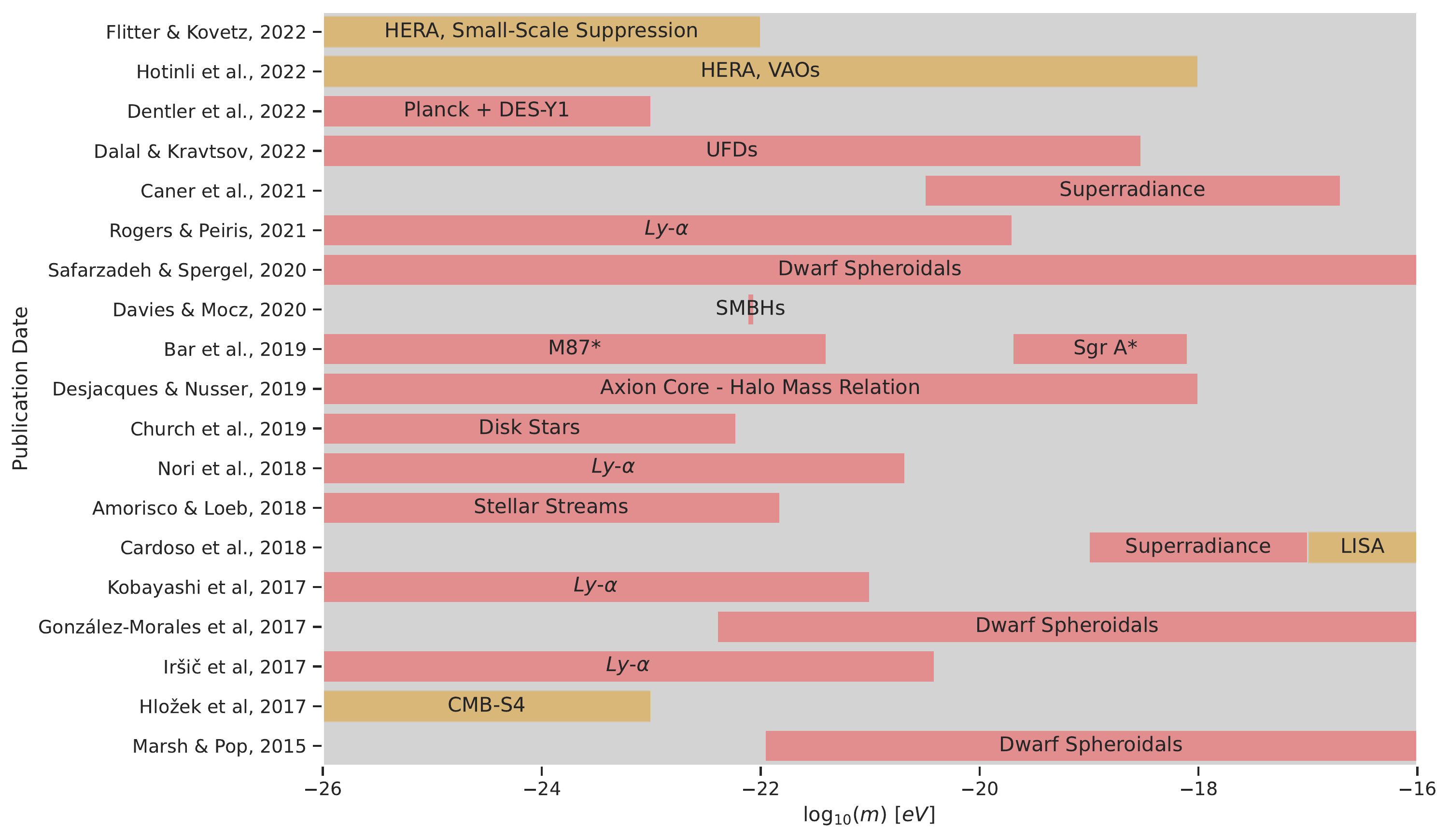}
\caption{Recent particle mass constraints for warm dark matter (WDM) / FDM from astrophysical observations in the FDM window of $10^{-26}$ eV $<m<10^{-16}$ eV. The red shading indicates a disfavored parameter space, though not necessarily a $2\sigma$-constraint. The orange shading indicates forecasts for future observatories. Note that most constraints assume a pure WDM / FDM cosmology with a fixed DM particle mass, rather than DM cocktails or a distribution of DM particle masses. The resolution / fidelity of the simulations underlying some of the constraints differ widely. UFDs = ultra-faint dwarf galaxies; SMBHs = supermassive black holes.}
\label{f_mass_constraints}
\end{figure*}

CMB data assuming a $\Lambda$CDM cosmology gives rise to an anomalous weak lensing amplitude $A_{\text{lens}}$ \citep{Calabrese_2008} and is in tension with low-redshift observations of Type Ia supernovae, standardisable quasars and cosmic chronometers \citep{Colgain_2022}. Even worse, $\Lambda$CDM faces several small-scale crises such as the ``missing satellite problem'' \citep{MooreGhigna_1999} and the ``core-cusp problem'' \citep{Wyse_2008}: Galaxies predicted by CDM extend to much lower masses, well below the observed dwarf galaxies, with steeper, singular mass profiles. Dwarf spheroidal (dSph) galaxies in particular exhibit a surprising uniformity of their central masses, $M(< 300 \ \text{pc}) \sim 10^7 M_{\odot}$ \citep{Strigari2008}, and shallow density profiles \citep{Bullock_2017}. Cf. \cite{Perivolaropoulos_2021} for a review on $\Lambda$CDM challenges.\par

In this work, we focus on fuzzy dark matter \citep[FDM]{Hui_2017}. The constituent particles in this dark matter (DM) model are drawn from an entire family \citep{Arvanitaki_2010, Visinelli_2019} of ultra-light bosons in the mass range $10^{-33}-10^{-19}$ eV. The lower bound, $m_H \sim 10^{-33}$ eV, is the Hubble scale and is equivalent to saying that the Compton wavelength of FDM is smaller than the visible Universe, while the upper bound depends on what masses are to be considered ultra-light. The uncertainty principle that these quantum particles obey counters gravity below a particle mass-dependent Jeans scale or de Broglie wavelength $\lambda_{\text{dB}}(m)$ \citep{Khlopov_1985}. The largely redshift-insensitive comoving de Broglie wavelength $\lambda_{\text{dB},c} \sim (1+z)^{\frac{1}{4}}m^{-\frac{1}{2}}$ simultaneously suppresses small-scale structure and limits the central density of collapsed halos \citep{Schive_2014}. The masses of cosmologically interesting FDM are traditionally expected to lie around $m \sim 10^{-22} - 10^{-20}$ eV: A dwarf galaxy in a halo of mass $M_h \sim 4\times 10^9 M_{\odot}/h$ and a virial radius $R_{\text{vir}} \sim 40 \ \text{kpc}/h$ will have a virial velocity $\varv_{\text{vir}} \sim 20$ km/s, which corresponds to a de Broglie wavelength of $\lambda_{\text{dB}} \sim 7 \ \text{kpc}/h$ assuming $m \sim 5\times 10^{-22}$ eV.\par 

Since the interparticle separation becomes smaller than the de Broglie wavelength for a DM particle mass of $m < 30$ eV, the ensuing wave character of FDM warrants the alternative term \textit{wave dark matter} \citep{Hui_2017}. The wave nature is reflected in the formation of interference patterns and \textit{solitonic cores} in centres of halos. Such cores match the observed flat cores of dwarf galaxies very well, and together with the suppression of small-scale structure solve some small-scale problems of $\Lambda$CDM.\par

However, experimental observations provide increasingly stringent bounds on the admissible mass ranges of FDM. Here, we disregard laboratory searches for axion-like signatures such as those involving haloscopes and helioscopes, some of which might have non-negligible sensitivity to ultra-light axions, i.e. FDM \citep{Irastorza_2018, Marsh_2016}. Instead, we focus on particle mass constraints from astrophysical observations. In Fig. \ref{f_mass_constraints}, we collate a non-exhaustive list of recent constraints ordered by publication date. While the density profiles in the central regions of dwarf spheroidals can be explained with a FDM soliton core potential provided that $m \lesssim 1.1 \times 10^{-22}$ eV \citep{Marsh_2015}, Ly-$\alpha$ observations find a lower bound $m \gtrsim 3.8 \times 10^{-21}$ eV to have enough power on Mpc-scales in the Ly-$\alpha$ forest \citep{Irsic_2017}, constituting a \textit{Catch-22} problem. Recently, \cite{Safarzadeh_2020} showed that a pure FDM cosmology is inconsistent with dwarf spheroidals across the entire parameter space by at least $3\sigma$. More sophisticated models of reionisation are needed though to make some of these observational constraints more reliable: For instance, most forest analyses assume that the frequency-averaged amplitude of the ionising
background $J$ in the neutral hydrogen density $n_{\text{HI}} \propto (1+\delta_b)^2/(T^{0.7}J)$ has negligible spatial fluctuations \citep{Hui_2017}. Some FDM constraints might also be biased due to poor star formation modelling.\par

High-redshift observations are most promising to provide evidence for / against FDM. The main reason can be traced back to non-linear structure formation, which redistributes power between scales so as to make the low-redshift FDM matter power spectrum resemble the $\Lambda$CDM one. For instance, in $N$-body simulations of the $m \sim 1.6 \times 10^{-22}$ eV model, by $z\sim 1$ discrepancies between the respective power spectra larger than $10$\% exist only for scales $k>20 \ h/$cMpc $\sim 2k_{1/2}$ \citep{Viel_2012}, corresponding to about twice the half-mode scale $k_{1/2}$. As we will see in Section \ref{s_highz_stats}, FDM imprints on the internal properties of virialised halos persist for somewhat longer than suggested by the power spectrum alone.\par

\subsection{Shapes of Halos and Galaxies}
\label{ss_shapes_halos_gxs}
Since the accretion of matter can be clumpy and directional, halo growth is anisotropic, resulting in the formation of nonspherical triaxial halos, see e.g. \cite{JeesonDaniel_2011}. In $\Lambda$CDM theory, a \textit{blue-nugget} transition for low-mass galaxies at $z \sim 2-4$ was identified, a term introduced in \cite{Dekel_2014}. Before the transition, a DM-dominated central body causes the majority of galaxies to be aligned with their DM halos. Both the galaxy and its host DM halo are on average fairly elongated, though as we will see in Section \ref{ss_shape_stats}, their shapes strongly depend on halocentric distance. The transition itself is mediated by compaction events such as mergers, counter-rotating streams or violent disc instabilities. Once complete, a self-gravitating baryonic core emerges, which scatters small-pericentre orbits of DM and stars, leading to a more spherical configuration. The outer DM halo remains less affected by baryonic self-gravitation in the centre.\par 

Is this transition motivated by semi-analytical arguments confirmed by simulations? \cite{Chua_2019} and \cite{Tomassetti_2016} noted that the baryonic core emerges by virtue of radiative processes such as radiative heating and cooling, star formation, chemical evolution as well as strong supernova and AGN feedback. These allow a proper condensation of baryons into the centre of halos, scattering DM particles that approach the halo centre and modifying box orbits into rounder passages. Eventually, a high star-formation rate (SFR) around \textit{cosmic noon} ($z\sim 2$) and a lower rate of gas inflow eventually leads to an inside-out quenching process into a compact, passive \textit{red nugget}, the likely progenitor of the centre of an early-type galaxy today.\par 

Given certain assumptions concerning the distribution of the 2D axis ratio $Q$ for a sample of observed galaxies, one can use the inferred values of $Q$ to put constraints on the distribution of the 3D axis ratios. For instance, \cite{Chang_2013} and \cite{van_der_Wel_2014} assumed that the triaxiality and the edge-on ellipticity are Gaussian distributed and concluded that low-mass galaxies (i.e. $M_{\star} \leq 7\times 10^{9}$ M$_{\odot}/h$) are more elongated at $z\gtrsim 1$, while they are consistent with a population of flattened spheroids at the current epoch, hinting again at the blue-nugget transition. In a nutshell, the observational basis for this picture is becoming solid, both for the red-nugget phenomenon \citep{Whitaker_2012} and for their potential blue-nugget progenitors \citep{Tacchella_2015, Takeuchi_2015}. Since we are focusing on high redshifts of $z\gtrsim 3.4$, baryonic effects on the shapes of DM halos are insignificant, further justifying our use of $N$-body simulations.\par 

\subsection{Correlated Alignments of Shapes / Spin}
While the high-$z$ statistical properties of halo shapes can already hint at the nature of DM as we will see in Section \ref{ss_shape_stats}, we gain additional insight when considering alignment correlations between halos. The simplest approach is to consider geometrical alignment measures such as shape-position and shape-shape alignments, defined in Section \ref{ss_def_align}. While these are easy to define, their measurement can be challenging, as exemplified in \citet[CANDELS Collaboration]{Pandya_2019}. They assigned each of the observed galaxies probabilities of being intrinsically elongated, flattened or spheroidal by essentially interpolating the results of \cite{Zhang_2019}. Yet they did not detect significant geometrical alignment signals in CANDELS observations, possibly owing to spectroscopic incompleteness.\par 

More apt alignment measures can be motivated via the mechanisms underlying them: Correlated alignments are typically driven by stretching or compression of initially spherically collapsing mass distributions in some gravitational gradient. Alternatively, alignments can result from the mutual acquisition of angular momentum through tidal torquing of aspherical protogalactic mass distributions during galaxy formation, for instance due to the proximity of a large scale cosmic filament. On large scales, the former effect is well captured by the Linear Alignment Model \citep[LAM]{Hirata_2004} while the latter is described by tidal torquing theory \citep[TTT]{Doroshkevich_1970, White_1984}, a second-order effect. Starting in the 2000s, intrinsic alignment as found in the SDSS and WiggleZ DES datasets proved to be consistent with LAM and TTT predictions \citep{Mandelbaum_2006, Hirata_2007, Lee_2011, Mandelbaum_2011}. However, the assumptions of TTT appear to be violated at lower redshift as hinted at by both simulations \citep{Zjupa_2020} and observations \citep[KiDS+GAMA]{Johnston_2019}. Given that we are post-processing $N$-body simulations, we will calculate LAM best-fits in Section \ref{ss_ia} while ignoring TTT.\par

Recently, the focus of alignment studies has shifted to isolating the contaminating contribution of large-scale intrinsic alignments to the weak gravitational lensing signal by the large-scale structure. In fact, intrinsic alignments are estimated to be one of the most serious physical systematic effects to the cosmic shear signal in the new era of systematics-dominated cosmology. The magnitude of the impact of intrinsic alignment on the observed lensing spectrum is comparable to the changes in cosmology even for a deep, stage IV weak lensing survey \citep{Troxel_2015}. This next generation of weak lensing surveys comprising the \href{http://www.euclid-ec.org/}{Euclid} mission, the \href{https://www.lsst.org/}{Vera Rubin Observatory} and the \href{https://roman.gsfc.nasa.gov/}{Nancy Roman Space Telescope} will all produce unprecedented weak lensing measurements in the coming decades and are all relying on proper alignment modelling. Understanding how FDM modifies intrinsic alignment correlations is an important ingredient in such models.

\subsection{Outline}
We focus on intermediate scales of $k \sim 2 - 18 \ h/$Mpc to study halo density profiles, shapes and alignments at high redshifts of $z > 3.4$ using a suite of $N$-body simulations of CDM and FDM-like scenarios. The paper is laid out as follows: In Section \ref{s_sims}, we describe our large-scale simulations. We present relevant definitions and the post-processing tools in Section \ref{s_methods}. Results for the high-$z$ statistics of FDM-like halos are given in Section \ref{s_highz_stats}. A summary follows in Section \ref{s_conclusions}. The Supplementary Materials are dedicated to initial conditions and pre-initial conditions in simulations with a power spectrum cutoff, to convergence tests and to the impact of quantum pressure on profiles of density and shape.

\section{CDM and \texorpdfstring{\lowercase{c}}{}FDM Simulations}
\label{s_sims}
We search for imprints of a small-scale cutoff in the initial DM power spectrum using cosmological $N$-body simulations performed with the state-of-the-art moving-mesh code \scshape{Arepo}  \normalfont \citep{Arepo_2010}. Gravitational forces are computed using a Tree-PM method where long-range forces are calculated on a particle mesh and the short-range forces are calculated using a tree-like hierarchical multipole expansion scheme.\par

Bona fide FDM simulations are much more challenging than CDM simulations as the oscillations with highest frequency $\omega \propto m^{-1}\lambda_{\text{dB}}^{-2}$ occur in the densest regions, requiring very fine temporal resolution even for moderate spatial resolution. For instance, a de-Broglie wavelength of $\lambda_{\text{dB}} = 0.6$ kpc$/h$ and velocity of $\varv=200$ km/s translate to an oscillation timescale of $\tau_{\text{osc}} = 2\pi/\omega \sim 3\times 10^6$ yrs$/h$. In this work, we use a widely adopted approximation for FDM which we shall call classical FDM (cFDM). As opposed to a bona fide superfluid, cFDM approximates FDM as a classical collisionless fluid, governed by the Vlasov-Poisson equation, but with FDM initial conditions. The initial small-scale suppression in the power spectrum is modelled using \scshape{AxionCamb} \normalfont  \citep{Hlozek_2015}. We will refer to this exponential-like suppression as a cutoff. By construction, cFDM ignores dynamical effects of the quantum pressure that are apparent on small scales \citep{Mocz_2020, Mocz_2019, May_2021}. Note that cFDM particles do not possess thermal velocities, i.e. they are not a valid WDM implementation.\par 

Why is cFDM a valid approximation? The scales of interest in this work are the intermediate ones of $k \sim 2 - 18 \ h/$Mpc, corresponding to halo virial masses $M_h = 4\pi (\pi/k)^3 \rho_m/3$ of $M_h \sim 2\times 10^9 - 10^{12} \ M_{\odot}/h$. On these scales, quantum pressure is not very significant. The reason is that the absolute fractional difference between growth rates in FDM vs cFDM does not exceed $10$\% for particle masses\footnote{We will use the FDM particle mass $m_{\text{FDM}}$ and the cFDM particle mass $m_{\text{cFDM}}$ interchangeably and use a shorthand notation, $m=m_{\text{FDM}}=m_{\text{cFDM}}$.} around $m\sim 10^{-22}$ eV and mass scales around $M \sim 2\times 10^9 \ M_{\odot}/h$ \citep{Corasaniti_2017, Schive_2016}. In fact, this fractional difference is less than $5$\% for mass scales beyond $M \sim 4\times 10^9 \ M_{\odot}/h$, representing the vast majority of the halos in our inventory, described in Section \ref{ss_halo_inventory}. Note that these estimates are based on the solutions to the linearised governing equations (Schr\"odinger-Poisson vs Vlasov-Poisson) extrapolated down to $z_{\text{end}} \sim 4$ in an EdS-like Universe without baryons, as in this work.\par

The $N$-body suite offers a much cleaner platform to assess the imprints of a cutoff in the initial power spectrum, as in full hydrodynamical runs subgrid baryonic physics uncertainties would make it challenging to disentangle the resolution effects that are due to baryonic physics from e.g. the resolution convergence of the iterative shape procedure outlined in Section \ref{ss_shape_finding}. For instance, \cite{Chua_2019} finds poor shape convergence in \scshape{Illustris}  \normalfont simulations that can be traced back to the under-prediction of galaxy stellar mass and galaxy formation efficiency at lower resolutions \citep{Vogelsberger_2013}.\par 

We use cosmological volumes with two different box side lengths, $L_{\text{box}} = 10$ and $40$ cMpc$/h$, for each of three DM resolutions, $N=256^3$, $512^3$, and $1024^3$. This set of simulation parameters best balances the competing demands of high resolution (for convergence in e.g. halo shapes) and large volume (to obtain accurate statistical distributions). It also allows to test convergence both in statistical distributions and with respect to mass resolution. Our $N$-body simulations have initial conditions generated at $z=127$ using \scshape{NGenIC} \normalfont \citep{Springel_2005}, an initial conditions code which relies on the Zeldovich approximation. The cosmological boxes are evolved until $z=3.4$. We run simulations over a range of bosonic particle masses, $m = 10^{-22} \ \text{eV}, 7\times 10^{-22} \ \text{eV}, 2\times 10^{-21} \ \text{eV}$. The fixed comoving Plummer-equivalent gravitational softening length is set to $\epsilon = 0.19$ ckpc/h. We adopt a Planck-like cosmology with $\Omega_m = 0.3089$, $\Omega_{\Lambda} = 0.6911$, $h=0.6774$, $\sigma_8 = 0.81$ (and $n_s = 0.9665$ for the $N$-body simulations of $\Lambda$CDM), cf. \cite{Planck_2015}. See Table \ref{t_sim_overview} for an overview of the $N$-body simulations.\par 

\renewcommand{\arraystretch}{1.6}
\begin{table*}
	\centering
    \begin{tabular}{c | c | c | c }
     \hline
     Simulation Type & Box Side Length [cMpc/h] & DM particles & $m_{\text{DM}}$ [$10^6 M_{\odot}$] \\ \hline
     \hline
     $\Lambda$CDM & $10$ & $256^3$ / $512^3$ / $1024^3$ & $5.11$ / $0.64$ / $0.08$\\ \hline
     $\Lambda$CDM & $40$ & $256^3$ / $512^3$ / $1024^3$ & $327$ / $40.9$ / $5.11$\\ \hline
     cFDM, $m=2\times 10^{-21}$ eV & $10$ & $256^3$ / $512^3$ / $1024^3$ & $5.11$ / $0.64$ / $0.08$\\ \hline
     cFDM, $m=2\times 10^{-21}$ eV & $40$ & $256^3$ / $512^3$ / $1024^3$ & $327$ / $40.9$ / $5.11$\\ \hline
     cFDM, $m=7\times 10^{-22}$ eV & $10$ & $256^3$ / $512^3$ / $1024^3$ & $5.11$ / $0.64$ / $0.08$\\ \hline
     cFDM, $m=7\times 10^{-22}$ eV & $40$ & $256^3$ / $512^3$ / $1024^3$ & $327$ / $40.9$ / $5.11$\\ \hline
     cFDM, $m=10^{-22}$ eV & $10$ & $256^3$ / $512^3$ / $1024^3$ & $5.11$ / $0.64$ / $0.08$\\ \hline
     cFDM, $m=10^{-22}$ eV & $40$ & $256^3$ / $512^3$ / $1024^3$ & $327$ / $40.9$ / $5.11$\\ \hline
    \end{tabular}
    \caption{Overview of the $N$-body simulations and the parameters used: (1) simulation type, including the particle mass $m$ in case of cFDM; (2) side length of simulation box; (3) number of DM resolution elements; (4) mass per DM resolution element. While the $10$ cMpc/h boxes are used for convergence tests, the bulk of this work analyses the $40$ cMpc/h boxes.}
    \label{t_sim_overview}
\end{table*}
\renewcommand{\arraystretch}{1}

Substructure is identified using the {\fontfamily{cmtt}\selectfont SUBFIND} algorithm explained in \cite{Springel_2001_2}. After identifying DM halos (called \textit{groups}) using the friends-of-friends ({\fontfamily{cmtt}\selectfont FoF}) algorithm with a standard linking length of $b = 0.2 \times \text{mean inter-particle separation}$, the algorithm searches for overdense regions inside each {\fontfamily{cmtt}\selectfont FoF} group and prunes them according to a gravitational boundedness criterion. In the case of cFDM, {\fontfamily{cmtt}\selectfont SUBFIND} identifies many particles beyond the virial radius $R_{\text{vir}}$ as part of the halo if they are gravitionally bound to it, providing an opportunity to study the ambient cosmic environment of halos in Section \ref{ss_shape_stats}.\par

\section{Post-Processing Tools}
\label{s_methods}

\subsection{Shape Determination Algorithm}
\label{ss_shape_finding}
Ideally, one would like to find the shapes of isodensity surfaces for a given 3D particle distribution. With only $\lesssim \mathcal{O}(10^4)$ particles, however, the next-best choice is to assume that the distribution can be reasonably fit by an ellipsoid, see \cite{Zemp_2011}. The axis ratios can be computed from the unweighted shape tensor $q_{ij}$, which is the mode-centred second moment of the mass distribution:
\begin{equation}
q_{ij} = \frac{1}{\sum_k m_k} \sum_k m_k r^{\text{mode}}_{k,i}r^{\text{mode}}_{k,j}.
\label{e_shape_tensor}
\end{equation}
Here, $m_k$ is the mass of the $k$-th particle, and $r^{\text{mode}}_{k,i}$ is the $i$-th component of its position vector with respect to the densest location within the cloud, i.e. mode. We have compared the potential centre definition with the result of the `shrinking sphere' algorithm that we use to find the densest location following \cite{Power_2003}, and have found very good agreement. The overall shape could be described by the axis ratios
\begin{equation}
q \coloneqq \frac{b}{a}, \ \ s \coloneqq \frac{c}{a},
\end{equation}
where $a, b$ and $c$ are the eigenvalues of $q_{ij}$ corresponding to the major, intermediate and minor axes, respectively. The ratio of the minor-to-major axis $s$ has traditionally been used as a canonical measure of the sphericity. A perfectly spherical cloud has $q=s=1$. The triaxiality parameter, first defined by \cite{Franx_1991} as
\begin{equation}
T \coloneqq \frac{1-q^2}{1-s^2},
\end{equation}
measures the prolateness / oblateness of a halo. $T=1$ describes a completely prolate (colloquially \textit{elongated}) halo, while $T = 0$ describes a completely oblate (colloquially \textit{flattened}) halo. In practice, halos with $T > 0.67$ are considered prolate while those with $T < 0.33$ are dubbed oblate. Halos with $0.33 < T < 0.67$ are said to be triaxial. \par

Note that $q$ and $s$ are in fact a function of distance from the centre of the particle cloud. A proper distance measure to choose is the ellipsoidal radius
\begin{equation}
r_{\text{ell}} = \sqrt{x_{\text{pf}}^2+\frac{y_{\text{pf}}^2}{(b/a)^2}+\frac{z_{\text{pf}}^2}{(c/a)^2}},
\end{equation}
measured from the mode. Here, $(x_{\text{pf}},y_{\text{pf}},z_{\text{pf}})$ are the coordinates of the particle in the eigenvector coordinate system of the ellipsoid (also called principal frame), i.e. $r_{\text{ell}}$ corresponds to the semi-major axis $a$ of the ellipsoidal surface through that particle.\par

We use an iterative process \citep{Katz_1991, Dubinski_1991} to compute the axis ratios $q(r_{\text{ell}})$ and $s(r_{\text{ell}})$. First, we discretise the major axis into logarithmically spaced radial bins. At a given semi-major axis $r_{\text{ell}} \equiv a$, we start by considering particles inside a sphere of radius $a$. Next, the shape tensor $q_{ij}(r_{\text{ell}})$, Eq. \eqref{e_shape_tensor}, is diagonalised which yields the eigenvectors and eigenvalues at distance $r_{\text{ell}}$. The eigenvectors give the directions of the semi-principal axes. As opposed to the differential version of this algorithm, the geometrical meaning of the eigenvalues remains somewhat uncertain and dependent on the mass density profile \citep{Zemp_2011}. It is generally assumed though that the eigenvalues of $q_{ij}$ are still proportional to the semi-major axes squared. With this approximation, we calculate the axis ratios
\begin{equation}
q = \sqrt{\frac{q_{yy}}{q_{xx}}}, \ \ s = \sqrt{\frac{q_{zz}}{q_{xx}}},
\end{equation}
where $q_{zz} < q_{yy} < q_{xx}$ are the principal components of the tensor. Now, we keep the length of the semi-major axis fixed - though the orientation can change - and calculate $q_{ij}$ again by summing over all particles within the new, deformed integration volume (= ellipsoid) with semi-major axis $a = r_{\text{ell}}$ and axis ratios $q$ and $s$. In other words, particle $(x_{\text{pf}}, y_{\text{pf}}, z_{\text{pf}})$ enters the summation in Eq. \eqref{e_shape_tensor} if
\begin{equation}
\sqrt{x_{\text{pf}}^2+\frac{y_{\text{pf}}^2}{q^2}+\frac{z_{\text{pf}}^2}{s^2}} \leq a.
\label{e_condition_retain_particle}
\end{equation}
This iteration is repeated until convergence is reached. As a convergence criterion, we typically require that the fractional difference between two iteration steps in both axis ratios is smaller than $10^{-2}$.\par 

Even if the particle resolution exceeds $\mathcal{O}(10^4)$ particles, we refrain from running the differential version of this algorithm which replaces ellipsoids with ellipsoidal shells of width $\Delta a$ \citep{Zemp_2011}. While some of our halos of the intermediate- and high-mass end would satisfy this criterion for the $1024^3$ resolution runs, many shells at large ellipsoidal radii struggle to converge with the differential version.\par 

It is instructive to estimate the convergence radius $r_{\text{conv}}$ beyond which we deem our shape calculations reliable. \cite{Power_2003} demonstrated that deviations from convergence depend (for appropriate choices of other numerical parameters) solely on the number of particles, and scale roughly with the collisional `relaxation' time, $t_{\text{relax}}$. The latter quantifies after how much time the actual velocity of test particles significantly differs from the velocity that they would have had if the mass of the other particles were smoothly distributed, rather than finitely-sampled. Since $t_{\text{relax}}(r)$ scales roughly like the enclosed number of particles times the local orbital time-scale, one often expresses $t_{\text{relax}}$ in units of the circular orbit time-scale $t_{\text{circ}}$ at $R_{200}$. According to $t_{\text{circ}}(R_{200})= 2\pi R_{200}/\varv_{R_{200}}=\sqrt{3\pi/(G\bar{\rho})} = \sqrt{3\pi/(200G\rho_{\text{crit}}\Omega_m)}$, the latter is of the order of the age of the Universe. The rescaled relaxation time may then be written as
\begin{equation}
\kappa(r) = \frac{t_{\text{relax}}(r)}{t_{\text{circ}}(R_{200})} = \frac{\sqrt{200}}{8} \frac{N(r)}{\ln N(r)}\left[\frac{\bar{\rho}(r)}{\rho_{\text{crit}}}\right]^{-1/2},
\label{e_def_kappa}
\end{equation}
where $N(r)$ is the enclosed number of particles and $\bar{\rho}(r)$ is the mean enclosed density within $r$. While \cite{Power_2003} found that to achieve deviations of no more than $10$\% in the local density profile requires $\kappa \sim 1$, \cite{Vera_2011} found that $\kappa = 7$ marked the convergence radius where the axis ratios of individual relaxed halos agreed in different resolution runs of a quiescent halo in the Aquarius simulations of Milky Way-mass dark matter halos. In other words, as our inner convergence radius $r_{\rho,\text{conv}}$ for density profiles we thus adopt the halocentric radius $r_{\rho,\text{conv}} = \langle r \rangle$, typically averaged over a mass bin, that satisfies $\kappa(r_{\rho,\text{conv}}) = 1$, and analogously for shape profiles using the ellipsoidal radius.\par 

Predicting halo shapes around the virial radius $R_{\text{vir}}$ is of importance due to the success of the spherical and ellipsoidal collapse models. However, there is also considerable interest in studying halo shapes at radii that correspond to the outskirts of the spiral galaxies which they may host, including our own Milky Way. We will thus put some focus on the specific radius of $R_{15} = 0.15 \ R_{\text{vir}}$. The choice of $R_{15}$ is motivated by observational measurements of the properties of the Milky Way halo, which using tidally disrupted dwarf satellites have been performed out to $40$ kpc$/h$ from the galactic centre \citep{Law_2010}. 

\subsection{Alignment Correlations}
\label{ss_def_align}
Following \cite[CANDELS Collaboration]{Pandya_2019}, we consider two distinct types of alignment: \textit{shape-position} alignments and \textit{shape-shape} alignments. As for the former, we are interested in how the major axis of a particle distribution relates to the vector pointing at the mode of another particle distribution in the sample. As for the latter measure, we quantify how much the major axes of two particle distributions deviate from one another.\par 

We compute the alignment angles trivially using
\begin{equation}
|\cos \theta | = \frac{|\mathbf{v}_1\cdot \mathbf{v}_2|}{\lVert \mathbf{v}_1 \rVert \ \lVert \mathbf{v}_2 \rVert}
\end{equation}
where $\theta$ is the alignment angle between the two vectors $\mathbf{v}_1$ and $\mathbf{v}_2$, and the absolute value is taken since shapes in the axisymmetric ellipsoidal approximation are spin-$2$ objects. For shape-position alignments, $\mathbf{v}_1$ is the major axis of one particle group while $\mathbf{v}_2$ is the position difference vector connecting the two modes. For shape-shape alignments, both $\mathbf{v}_1$ and $\mathbf{v}_2$ are major axes.\par 

Central subhalos can be identified using {\fontfamily{cmtt}\selectfont SUBFIND}: The subhalo in each {\fontfamily{cmtt}\selectfont FoF} group with the lowest-potential resolution element is classified as central. They are expected to exert a continuous gravitational torque on non-centrals, colloquially called satellites, aligning them partially with the halo. This is exemplified by the Halo Alignment Model by \cite{Schneider_2010}. Since we are interested in alignments mediated by the larg-scale structure, we discard satellites in our alignment studies. In fact, we discard satellites throughout this entire work, including for density and shape profile estimations, all of which they would bias otherwise.\par 

\subsection{Halo Inventory}
\label{ss_halo_inventory}
As a minimum halo resolution we reject all halos consisting of fewer than $400$ DM resolution elements, i.e. $M_{\text{min}} = 400 \times m_{\mathrm{DM}} = 2.0 \times 10^9 M_{\odot}/h$ for the runs with $1024^3$ resolution, $L_{\text{box}} = 40$ cMpc$/h$. This resolution floor is sufficient to discard the vast majority of small-mass clumps that stem from the artificial fragmentation of filaments \citep{Lovell_2014}, even for the $m=10^{-22}$ eV halo sample. We conclude this by comparing $M_{\text{min}}$ to the location of the artificial small-mass upturn in the inferred halo mass function (not shown). At redshift $z=4.38$, this leaves us with $\sim 46000$ halos for CDM, $\sim 39000$ halos for cFDM with $m=2\times 10^{-21}$ eV, $\sim 32000$ halos for cFDM with $m=7\times 10^{-22}$ eV and $\sim 5000$ halos for cFDM with $m=10^{-22}$ eV. The ratios of these numbers are in good agreement with the numbers obtained when extrapolating the \cite{Schneider_2012} WDM-to-CDM recalibration fit
\begin{equation}
\frac{\mathrm{d}n^{\text{WDM}}/\mathrm{d}M}{\mathrm{d}n^{\text{CDM}}/\mathrm{d}M}(M) = \left(1+\frac{M_{1/2}}{M}\right)^{-\beta}
\end{equation}
with $\beta = 1.16$ and half-mode mass $M_{1/2}$ (see Eq. \eqref{e_half_mode}) to higher redshifts. Here, we use the half-mode matching formula
\begin{equation}
m_X = 0.84 \left(\frac{m}{10^{-22}\text{eV}}\right)^{0.39} \ \text{keV}
\label{e_half_mode_matching}
\end{equation}
to translate from a cFDM particle mass $m$ to a thermal relic WDM mass $m_X$ \citep{Marsh_2016}. When considering mass density profiles, we impose additional relaxedness conditions outlined in Section \ref{ss_rho_profs} which reduce the number of halos by a factor of $\sim 0.2$. Finally, for the construction of median shape profiles we additionally discard halos whose shape determination fails at the virial radius $R_{\text{vir}}$, leading to a reduction in the number of halos by a factor of $\sim 0.25$.\par

\section{Density, Shape and Alignment Statistics}
\label{s_highz_stats}
To assess the high-$z$ statistics of dark matter halos in the cFDM cosmologies, we investigate four statistical measures of the large-scale structure in order of complexity.
\subsection{Inferred Power Spectra}
The simplest such measure is the two-point statistics of the DM density field, which is traced by DM resolution elements in the \scshape{Arepo} \normalfont simulations. The theoretical \scshape{AxionCamb} \normalfont power spectra at $z=127$ drop by many orders of magnitudes at high wavenumbers in an approximately exponential fashion, accompanied by various axion wiggles. The resolvability of the cutoff with a finite number of DM resolution elements is an indicator of the quality of the initial conditions. We find that the inferred power spectra from the simulations replicate the cutoff down to very low power spectra magnitudes of $P(k) \sim 10^{-13}$ $h^{-3}$Mpc${}^3$. For a cFDM cosmology with $m=10^{-22}$ eV, the latter value is attained at a spatial scale of $k\sim 30$ $h/$Mpc.\par 

\begin{figure}
\hspace{-0.3cm}
\includegraphics[scale=0.54]{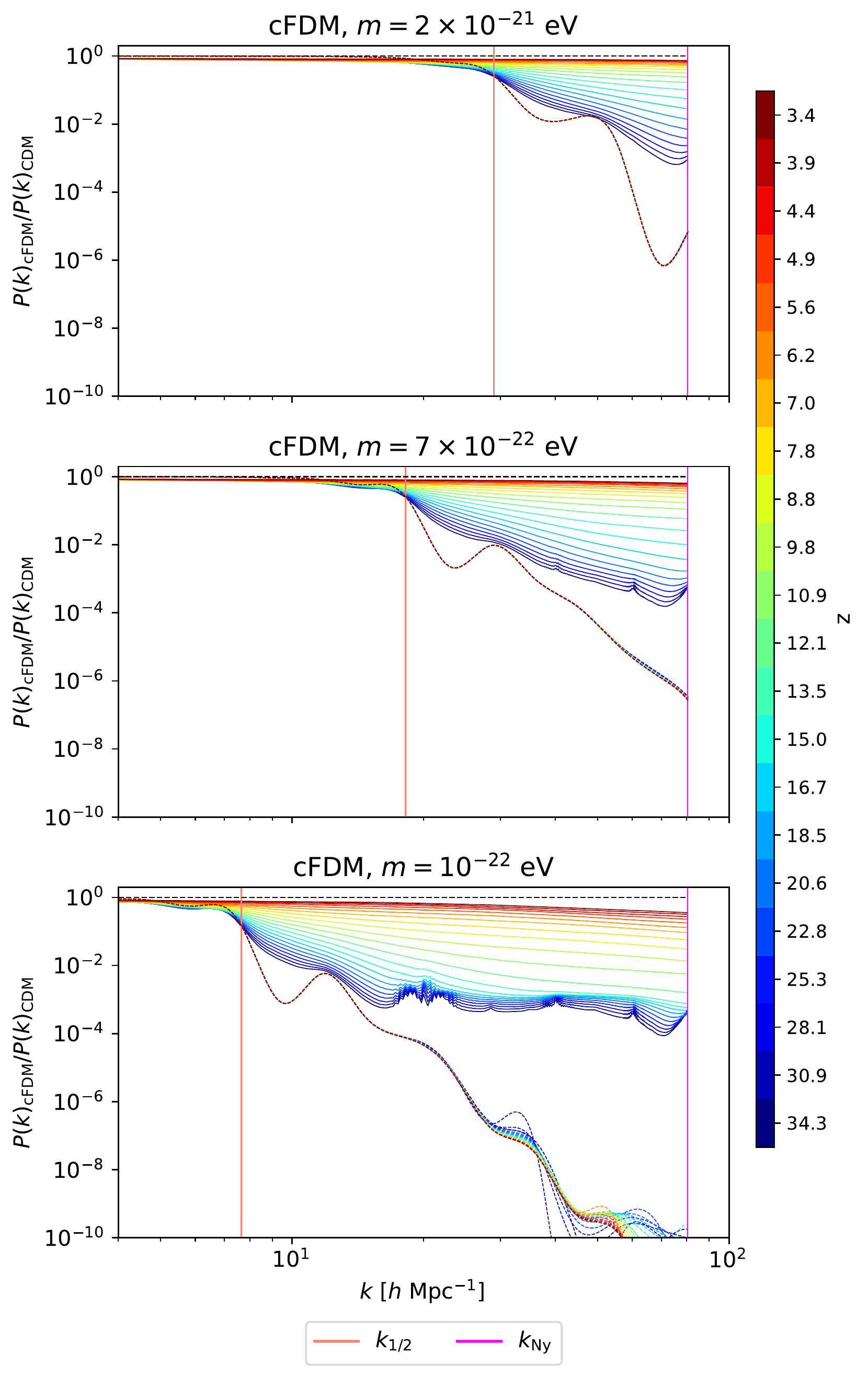}
\caption{Evolved power spectrum ratios $P(k)_{\mathrm{cFDM}}/P(k)_{\mathrm{CDM}}$ of the comoving DM density field, color-coded by redshift. Results are shown for $N$-body runs with $1024^3$ resolution and $L_{\text{box}} = 40$ cMpc$/h$. The vertical lines mark the FDM half-mode scale $k_{1/2}$ (orange) and the Nyquist frequency $k_{\text{Ny}}$ (magenta). The DM model is indicated on top of each panel and the redshift is specified by the colorbar labelling. Ratios of linear power spectra as obtained with \scshape{AxionCamb} \normalfont at the corresponding redshifts are traced by the dashed curves.}
\label{f_pspectra}
\end{figure}

The ratio of cFDM to CDM power spectra for different axion masses are shown in Fig. \ref{f_pspectra} for a wide range of redshifts\footnote{The power spectra were obtained using \textit{nbodykit}, a versatile open-source Python package \citep{Hand_2018}.}. To obtain such high-$k$ power spectrum estimation fidelity, we specified a non-canonical Piecewise Cubic Spline (PCS) mass assignment scheme for the kernel of the window function $W(\mathbf{x})$ to obtain the density field on a mesh of resolution $1024^3$. PCS is a third-order window function, i.e. there are $p=3+1$ grid points per dimension to which each particle is assigned \citep{Sefusatti_2016}. As a comparison, the widely adopted cloud-in-cell (CIC, cf. \cite{Hockney_1981}) mass assignment scheme is a first-order window function. We compensate for the window function $W(\mathbf{x})$ by trivially deconvolving the kernel in Fourier space.\par

The color-coding in Fig. \ref{f_pspectra} reflects the logarithmically spaced scalefactors $a$ that were chosen for the snapshots. Constant spacing in the logarithm of the scalefactor has desirable properties when considering galaxy populations and cosmology, see \cite{Baldry_2018}. Since the differences between two scalefactors thus translate to approximately (exactly in Einstein-de-Sitter space) constant comoving distances, the power spectra amplitudes undergo approximately constant multiplicative shifts as time evolves, at least in the regime of linear structure formation. We show curves for $34.3\geq z\geq 3.4$.

The half-mode scale (vertical orange line in Fig. \ref{f_pspectra}) is illustrative to consider since the cutoff occurs over a range of scales, and following \cite{Marsh_2016} the definition we adhere to is
\begin{equation}
T(k_{1/2}) = \left(\frac{P_{\text{lin}}^{\text{FDM}}(k_{1/2},z)}{P_{\text{lin}}^{\text{CDM}}(k_{1/2},z)}\right)^{\frac{1}{2}} = \frac{1}{2}, \ \forall z \geq 0,
\label{e_half_mode}
\end{equation}
where $T(k)$ is the redshift-independent FDM-to-CDM linear transfer function ratio\footnote{Note that a redshift-independent $T(k)$ is only a good approximation for $m\gtrsim 10^{-24}$ eV \citep{Marsh_2016}.}.\par

The linear theory predictions in Fig. \ref{f_pspectra} are obtained with \scshape{AxionCamb} \normalfont  and thus bona fide linear FDM theory. The fact that the linear theory power spectrum ratios of FDM to CDM do not all overlap is indicative of scale-dependent growth, a feature of FDM even when its governing equations (Schr\"odinger-Poisson) are linearised \citep{Marsh_2016}. For CDM and cFDM, linear growth always corresponds to scale-independent growth due to the properties of the Vlasov-Poisson system of equations. The non-linear predictions as obtained with the $N$-body code show that at high redshifts of $z\gtrsim 25$ and large scales both CDM and cFDM follow the standard linear evolution of the power spectrum $P(k,a) \propto a^2 P(k)$ with growing scale-factor $a$. This is indeed predicted for the matter-dominated era, with distinct modes evolving independently as a distribution of Gaussian random fields. It is the smaller scales for which this simple approximation breaks down first. Instead, non-linear structure formation successfully redistributes power to smaller scales.

For the cFDM cosmologies, we observe a wiggly pattern of power at high $k$ which goes beyond the mere axion wiggles and is thus indicative of numerical \textit{lattice features}. As the grid pre-initial condition (cf. Supplementary Materials) that we employ simply places $N^3$ particles onto the grid points of a three dimensional Cartesian lattice, the regular grid spacing introduces periodic signals into the pre-initial condition. The periodic signals get reinforced by gravity at high redshift while actual structure formation eventually overtakes the $N$-body discreteness effects in amplitude. The lattice features can persist in low-density regions of very low-resolution simulations even at redshift $z = 0$, cf. \cite{Wang_2007} and \cite{Liao_2018}. It is evident from Fig. \ref{f_pspectra} that the redshift $z_{\ast}$ at which lattice features become insignificant on \textit{all} resolvable scales is dependent on the particle mass $m$, or equivalently the half-mode scale $k_{1/2}$. More precisely, $z_{\ast}$ grows with increasing $m$. Amongst the $1024^3$ resolution, $L_{\text{box}} = 40$ cMpc$/h$ runs, lattice features are most profound for the $m=10^{-22}$ eV scenario for which they become unrecognisable on all resolvable scales around $z_{\ast}\sim 12$.\par

We admit that in our convergence tests in which we also employ simulations of resolution $256^3$ and $512^3$, some unavoidable lattice-induced small-scale power is still present, especially for the $256^3$ runs with $m=10^{-22}$ eV, $L_{\text{box}} = 10$ cMpc$/h$ and n.b. $L_{\text{box}} = 40$ cMpc$/h$. However, halo density and shape distributions and by extension shape alignments are robust against very low-magnitude small-scale power. The question of ideal initial loads for simulations with power spectra cutoffs is an intricate one. We try to answer this question in the Supplementary Materials, in favour of grid pre-initial conditions.\par

\subsection{Density Profiles}
\label{ss_rho_profs}

\begin{figure*}
\includegraphics[scale=0.6]{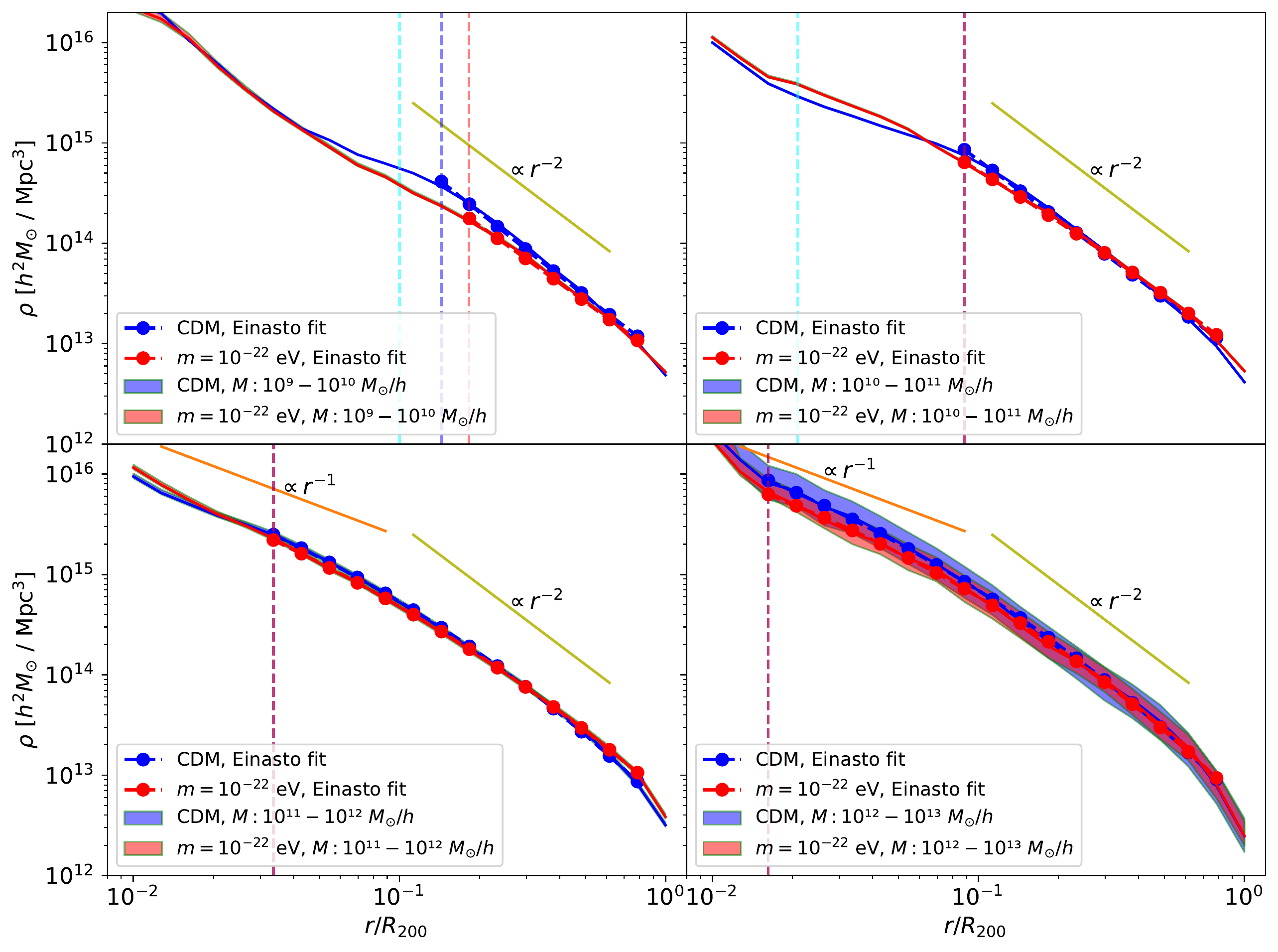}
\caption{Density profiles for $N$-body, $1024^3$ resolution, $L_{\text{box}} = 40$ cMpc$/h$ runs at redshift $z=4.38$, for cFDM with particle mass $m=10^{-22}$ eV (red), vs. CDM halos (blue). The mass bins span $10^9-10^{10}M_{\odot}/h$ (top left) all the way to $10^{12}-10^{13}M_{\odot}/h$ (bottom right). See legends for details. The shaded areas delineate the $25 - 75$th percentiles. Red and blue dashed verticals denote $\kappa = 1$ lower convergence radii, naturally moving towards smaller halocentric radii for higher-mass halos. The $\kappa = 1$ lines for CDM and cFDM overlap in most panels. The $r^{-1}$ and $r^{-2}$ power laws trace two characteristic regimes of an NFW-like profile. The cyan dashed verticals show the \citep{Schive_2014_2} estimates for the soliton core radius that one would find in bona fide FDM simulations. For the two higher-mass bins, the core radius is too small to be shown, $r_c/R_{200} = 4.5 \times 10^{-3}$ for $10^{11}-10^{12}M_{\odot}/h$ and $r_c/R_{200} = 1.1 \times 10^{-3}$ for $10^{12}-10^{13}M_{\odot}/h$.}
\label{f_rho}
\end{figure*}

Before we continue in Section \ref{ss_shape_stats} with shape statistics results, we first remind the reader of some well-known results for FDM-like halo density profiles and then generalise them to the new parameter space.\par

It is well established that quasi-equilibrium CDM halos have an approximately universal density profile that can be reproduced by rescaling the NFW profile \citep{Navarro_1996}
\begin{equation}
\frac{\rho(r)}{\rho_{\text{crit}}} = \frac{\delta_c}{(r/r_s)(1+r/r_s)^2}.
\end{equation}
The latter is fully determined by two parameters, the characteristic or scale radius $r_s \equiv r_{-2}$ at which the logarithmic slope has the isothermal value of $-2$, i.e. $\mathrm{d} \ln{\rho} / \mathrm{d} \ln r|_{r_{-2}} = -2$, and a characteristic overdensity $\delta_c$. This universality of spherically averaged density profiles holds over many orders of magnitudes in mass, cosmological parameters, magnitude and shape of the initial density fluctuation power spectrum and even in several modified gravity models \citep{Wang_2009, Hellwing_2013}. However, halo density profiles are better approximated with the Einasto profile \citep{Einasto_1965}
\begin{equation}
\ln\left(\frac{\rho_E(r)}{\rho_{-2}}\right) = -\frac{2}{\alpha}\left[\left(\frac{r}{r_{-2}}\right)^{\alpha}-1\right].
\label{e_def_einasto}
\end{equation}
While it is often reported that the third parameter is needed to more accurately capture the halo-to-halo variation in profile shape as well as its mass-dependence, this conclusion is not correct. For even if the third parameter $\alpha$ is fixed as a function of halo mass and redshift to around $\alpha \sim 0.17$, the Einasto profile provides a better fit to simulation results \citep{Gao_2008}.\par 

It has been found \citep{Gonzalez_2016} that low-redshift WDM halo profile shapes can be well accounted for by the NFW and Einasto profiles. We thus use the Einasto profile as our fitting model for cFDM cosmologies and assess the fidelity thereof a posteriori. We obtain $\rho(r)$ curves by brute-force binning of DM resolution elements into 20 mode-centred radial bins, equally spaced in $\log r$ spanning $-2.0 \leq \log (r/R_{200}) \leq 0.0$. We also compute the total enclosed mass $M(r)$ and mean inner density profiles $\bar{\rho}(r) = M(r)/(4/3)\pi r^3$ in order to compute $\kappa$-profiles according to Eq. \eqref{e_def_kappa}.\par 

We construct the $c(M,z)$ relation by largely following \cite{Ludlow_2016}: First, we remove substructure contamination by focusing on particles identified by {\fontfamily{cmtt}\selectfont SUBFIND} as part of central subhalos. We then fit the \textit{median} mode-centred density profiles after averaging over logarithmic mass bins\footnote{All of the mass bins in this work refer to {\fontfamily{cmtt}\selectfont FoF} masses.} of width $\Delta \log M = 0.2$. The advantage of this approach is that it smooths out any features unique to individual systems and dampens the influence of outliers. Best-fit Einasto profiles are determined by adjusting the three parameters of Eq. \eqref{e_def_einasto} in order to minimise a figure-of-merit defined as
\begin{equation}
\psi^2 = \frac{1}{N_{\text{bin}}} \sum_{i = 1}^{N_{\text{bin}}} \left[\ln \rho_i - \ln \rho_E(r_i; \rho_{-2}; r_{-2}; \alpha)\right]^2.
\end{equation}
Here, the number of radial bins is $N_{\text{bin}} < 20$ since we discard those that fall below the inner convergence radius estimate $r_{\rho,\text{conv}}$ given by $\kappa (r_{\rho,\text{conv}}) = 1$. This choice for $r_{\rho,\text{conv}}$ differs from the choice in \cite{Ludlow_2016}: They choose $r_{\rho,\text{conv}} = 3\epsilon$, where $\epsilon$ is the gravitational softening length. While both estimates for $r_{\rho,\text{conv}}$ were first introduced in \citep{Power_2003}, the latter is less robust since it constitutes a halo mass-independent and thus halo resolution-independent estimate. In addition, radial bins that exceed the outer limit of $0.8 \ R_{200}$ are also discarded in the fit since they may correspond to radii that are not fully relaxed \citep{Ludlow_2016}.\par

To account for smooth accretion, minor mergers and occasional major mergers with systems of comparable mass that drive individual halos out of quasi-equilibrium, we choose to discard unrelaxed halos by combining two conditions. The first is to request halos to have a normalised offset between the mode $\mathbf{r}^{\text{mode}}$ of the halo and its centre of mass $\mathbf{r}^{\text{CoM}}$ of no more than $d_{\text{off}} = |\mathbf{r}^{\text{mode}}-\mathbf{r}^{\text{CoM}}|/R_{200}< 0.07$. As mentioned in Section \ref{s_sims}, some cFDM halos are deeply embedded into nearby cosmic filaments, especially at high redshift. Consequently, we choose to evaluate $d_{\text{off}}$ by excluding DM resolution elements outside of $R_{200}$ to avoid filament-induced biases. The second condition we impose is that the virial parameter $\eta = (2K-S_p)/|W| < 1.35$. Here, $K$ is the kinetic energy of the halo, $W$ its potential energy and $S_p$ the surface pressure term
\begin{equation}
S_p = \oint \rho (\mathbf{x}\cdot\mathbf{v})(\mathbf{v}\cdot \mathrm{d}\mathbf{S}) = 4\pi R_{\text{vir}}^3\rho_{\text{vir}}\varv_r^2
\end{equation}
that follows from relaxing the assumption that the density at the halo boundary is zero \citep{Davis_2011, Klypin_2016}. The surface integral is taken over the surface of a sphere of virial radius $R_{\text{vir}}$, and the density $\rho_{\text{vir}}$ and radial velocity dispersion $\varv_r$ at the virial radius can be estimated from the halo density profile. The surface pressure correction is of considerable importance in the high redshift, rapid merging phase of the Universe, and upon its exclusion one would categorise many relaxed halos as unrelaxed \citep{Klypin_2016}. \par

We avoid using alternative relaxedness conditions such as the one based on the substructure mass fraction $f_{\text{sub}} = M_{\text{sub}}(<R_{200})/M_{200}$ \citep{Neto_2007}. It is often constrained to $f_{\text{sub}} < 0.1$, yet the inferred $f_{\text{sub}}$ is resolution-dependent and thus heavily underestimated for low-mass systems. \cite{Ludlow_2016} resorts to a dynamical age requirement to flag halos undergoing rapid accretion and equal-mass merging, a largely redundant condition in view of our imposed $\eta < 1.35$. In other words, our set of two conditions proves sufficient to remove the ``upturn'' in the concentration of high-mass halos reported by e.g. \cite{Klypin_2011}.\par

The results of the Einasto fitting procedure for four mass bins in the $N$-body, $1024^3$ resolution, $L_{\text{box}} = 40$ cMpc$/h$ runs at redshift $z=4.38$ are shown in Fig. \ref{f_rho}. The curves show the median density profile in the respective mass bin while the shaded area delineates the $25 - 75$th percentile, measuring the halo-to-halo variation. We find that both CDM and cFDM halos for the $m=10^{-22}$ eV scenario have density profiles that can be well fit with an Einasto profile. The red and blue dashed verticals demarcate the convergence radius $r_{\rho,\text{conv}}$ defined by $\kappa(r_{\rho,\text{conv}}) = 1$, as explained around Eq. \eqref{e_def_kappa}. As is evident from said equation, larger-mass halos have smaller convergence radii. In the plotted range of radii, their density profiles also do not exhibit spuriously large densities close to the halo centers, as opposed to small-mass halos (first row).\par 

The core radii of solitons that one would encounter in bona fide FDM simulations are given by cyan dashed verticals following \citep{Schive_2014_2}. The well-known trend of core radii growing with decreasing halo mass is recovered. In the mass bin $10^{11}-10^{12}M_{\odot}/h$, the soliton core radius $r_c/R_{200}\sim 4 \times 10^{-3}$ normalised by the median $R_{200}$ value is not shown and similarly for the mass bin $10^{12}-10^{13}M_{\odot}/h$ for which $r_c/R_{200}\sim 10^{-3}$. Even in the lowest mass bin of $10^{9}-10^{10}M_{\odot}/h$, the core radius is smaller than the convergence radius, demonstrating that we would not expect solitons to contaminate our density profile inferences at this level of resolution. In the Supplementary Materials, we show in detail how quantum pressure in FDM induces alterations of the halo density profile.\par

\begin{figure}
\hspace{-0.7cm}
\includegraphics[scale=0.7]{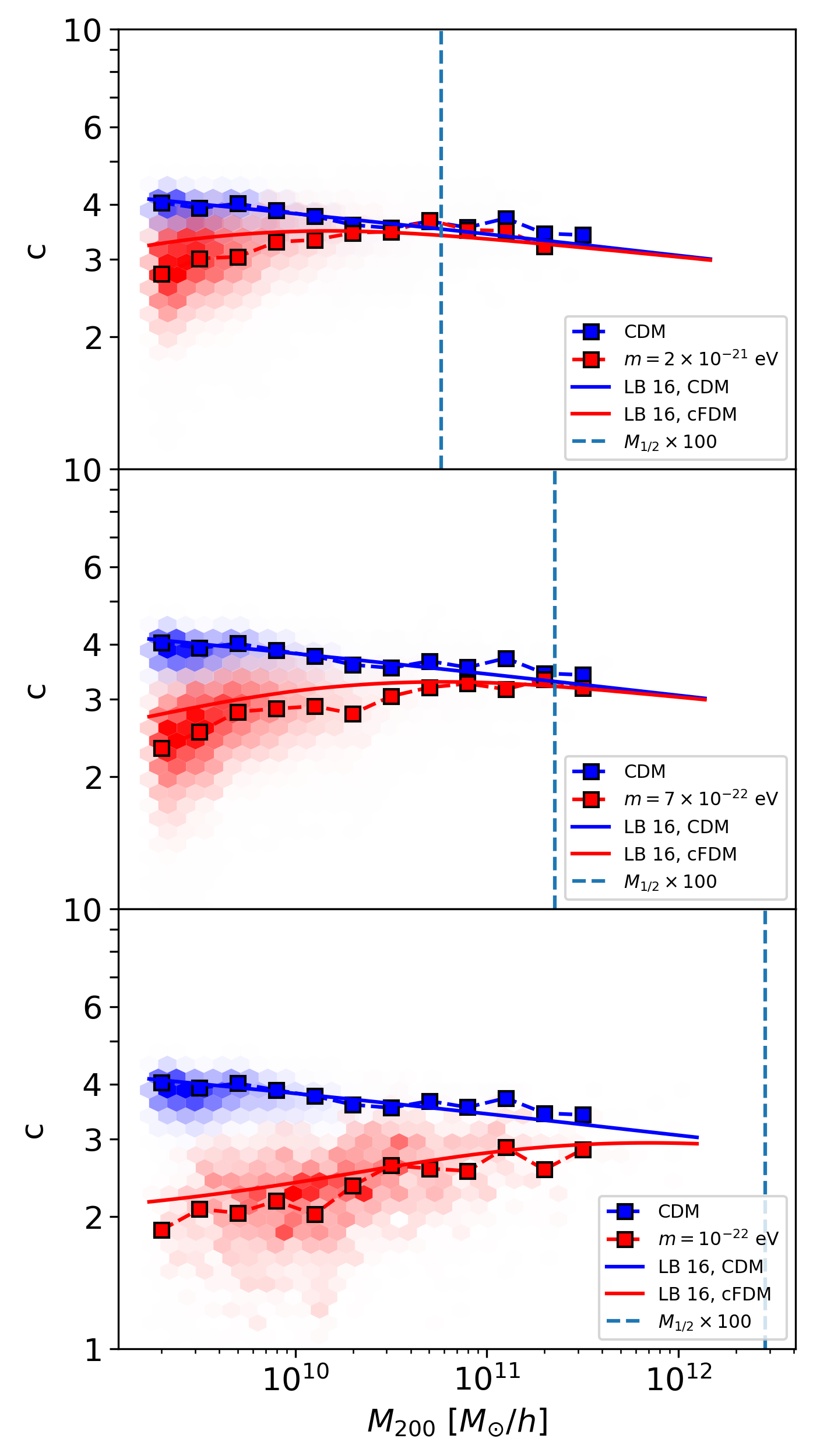}
\caption{Concentration-mass relation $c(M,z)$ in different cosmologies, for $N$-body, $1024^3$ resolution, $L_{\text{box}} = 40$ cMpc$/h$ runs at redshift $z=4.38$. We compare the CDM halos (blue) with cFDM halos for particle mass $m=2\times 10^{-21}$ eV (top), $m=7\times 10^{-22}$ eV (middle) and $m=10^{-22}$ eV (bottom). Square markers indicate median concentrations obtained from Einasto best-fits while the color of each hexbin is proportional to the number of halos whose Einasto concentration falls therein. `LB 16, CDM' and `LB 16, cFDM' show an analytical model prediction from \protect\cite{Ludlow_2016} for model parameters $f=0.02$ and $C=650$ (see text).}
\label{f_cVersusM}
\end{figure}

Finally, note that in CDM the median scale radius $r_{-2}$ migrates towards larger radii as the halo mass grows. This is of great importance. In fact, the major advantage of NFW and Einasto profiles is that their scale radius $r_s$, often called concentration radius, can be used to define the halo concentration
\begin{equation}
c \coloneqq \frac{R_{200}}{r_{-2}}
\end{equation}
as the ratio of $R_{200}$ to that of the scale radius. This definition can be extended straightforwardly to alternative DM scenarios such as WDM and FDM. As discussed by NFW for the case of CDM, $c$ and $M_{200}$ do not take on arbitrary values, but correlate in a way that reflects the mass-dependence of halo formation times \citep{Bullock_2001}. Earlier assembly corresponds to higher characteristic densities, reflecting the larger background density at that epoch.\par 

\begin{figure}
\vspace{0.15cm}
\includegraphics[scale=0.70]{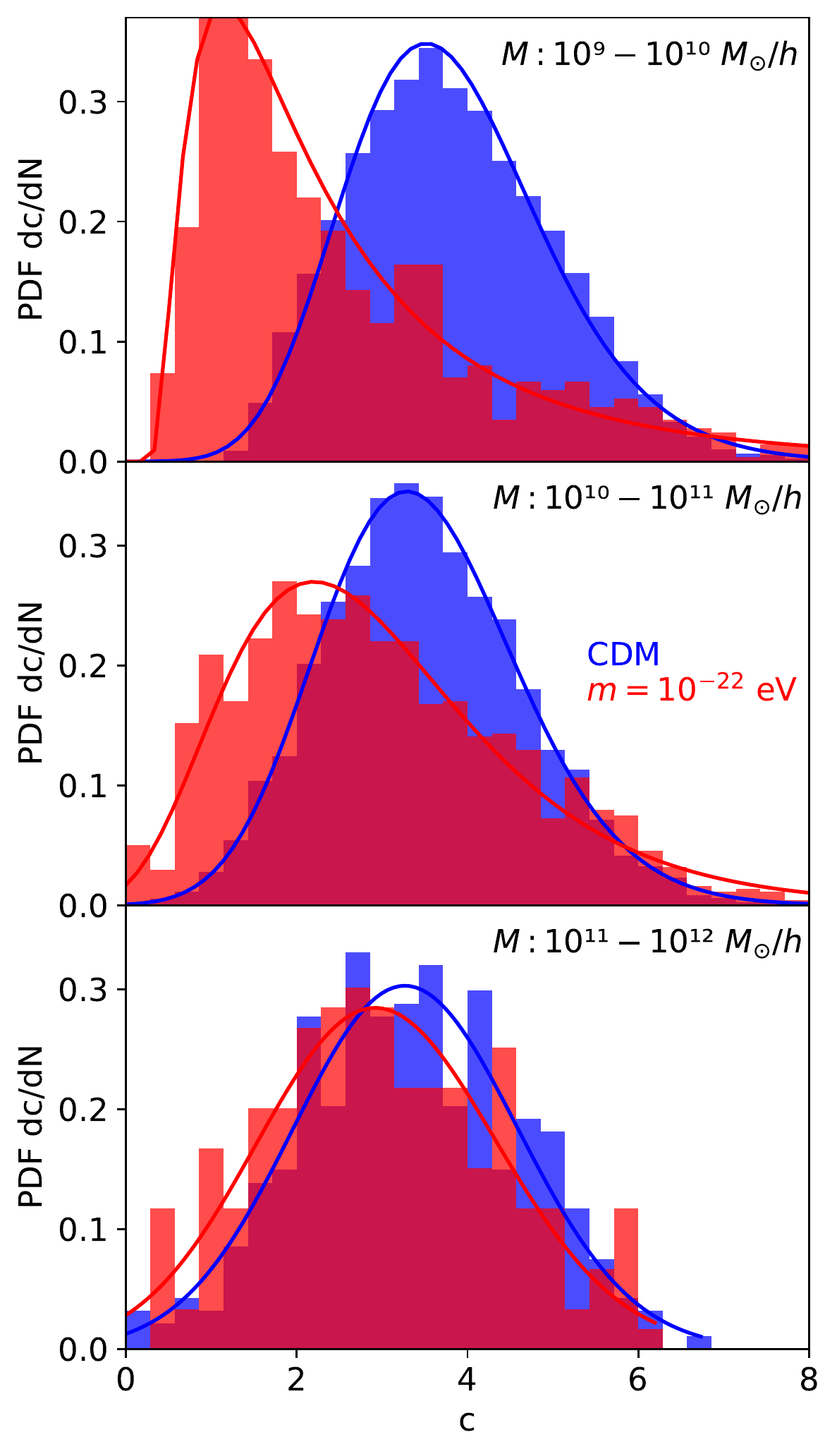}
\caption{Concentration probability distribution functions in different cosmologies, for $N$-body, $1024^3$ resolution, $L_{\text{box}} = 40$ cMpc$/h$ runs at redshift $z=4.38$. We compare CDM halos (blue) for mass bin $10^9-10^{10}M_{\odot}/h$ (top), $10^{10}-10^{11}M_{\odot}/h$ (middle) and $10^{11}-10^{12}M_{\odot}/h$ (bottom) with cFDM halos for particle mass $m=10^{-22}$ eV. Solid curves are lognormal best fits.}
\label{f_cPDF}
\end{figure}

\cite{Ludlow_2014} later showed that the concentration of a CDM halo can be inferred from the critical density of the Universe at a characteristic time along its mass accretion history. The insight was used to construct an analytic model for the mass-concentration-redshift relation $c(M,z)$ for CDM halos, reproducing the inferred relation from many CDM simulations. In \cite{Ludlow_2016}, the generalisation of the model has been introduced and applied to WDM, relying on the full `collapsed mass history' of a halo instead of its main progenitor. As such, the model is based on extended Press-Schechter theory, allowing to assess the dependence of halo concentrations on cosmological parameters and on the shape of the linear matter power spectrum alike. It is this model whose validity for cFDM and high redshifts we would like to comment on.\par 

Fig. \ref{f_cVersusM} shows parts of the $c(M,z)$ relation as inferred from the $N$-body, $1024^3$ resolution, $L_{\text{box}} = 40$ cMpc$/h$ runs at redshift $z=4.38$. Square markers indicate concentrations as obtained from Einasto best-fits in the respective mass bin, while the color of each hexbin is proportional to the number of halos whose concentration falls therein. The solid blue curve labelled `LB 16, CDM' is a broken power-law fitting formula for $c(M,z)$ in the Planck cosmology from \cite{Ludlow_2016} (Eq. C1). At fixed redshift $z$, the concentrations of CDM halos decrease monotonically with increasing halo mass, reflecting the lower background density at the epoch of their formation. At fixed mass, concentrations $c$ decrease monotonically with increasing redshift (not shown), indicative of the redshift evolution of the reference density, also called the \textit{pseudo-evolution} of mass \citep{Diemer_2013}.\par

For cFDM, the concentration-mass relation is non-monotonic. In solid red labelled `LB 16, cFDM', we add the model prediction from \cite{Ludlow_2016}. To maintain consistency, as input to the model we use linear FDM matter power spectra obtained with \scshape{AxionCamb} \normalfont  rather than the \cite{Viel_2005} parametrisation suitable for WDM. The model parameter $f$ that is used to define the halo collapse redshift is set to $f = 0.02$ while the proportionality factor $C$ relating mean inner halo densities to the critical density of the Universe at the collapse redshift is set to $C = 650$. We find that their model slightly overpredicts the cFDM concentrations at the small-mass end, while performing better for higher masses. At given $z$, concentrations peak at around two orders of magnitude above the half-mode mass scale $M_{1/2}$, with concentrations declining above and below this \textit{peak mass scale} $M_{\text{peak}} \sim 100 \times M_{1/2}$. This is a known result which has also been described by e.g. \cite{Bose_2015}. However, both the latter work and that of \cite{Ludlow_2016} focus on WDM simulations at low redshifts of $z\leq 3$ and equivalent cFDM particle masses of $3.3\times 10^{-21}$ eV and $4.5\times 10^{-22}$ eV. We extend the result to cFDM of lower particle mass of $m = 10^{-22}$ eV and to redshifts in the range $3.4 \leq z \leq 6.3$, beyond which our halo catalogues become too sparse and noisy to make statistically robust conclusions. We concede that for the $m = 1 \times 10^{-22}$ eV run (lower panel), our lack of high-mass halos makes the extrapolation towards $M_{\text{peak}}$ only tentative.\par

As in WDM, the non-monotonicity of $c(M,z)$ is due to the suppression of gravitational collapse below the half-mode scale, breaking the scale-invariance of the assembly process. It imprints a preferred scale on the mass accretion histories. While \cite{Ludlow_2016} apply their model of the $c(M,z)$ relation to WDM, here we show  that it also fares well for cFDM simulations with initial conditions based on \scshape{AxionCamb}\normalfont , reproducing the non-monotonicity and even the constancy of $M_{\text{peak}}$ with cosmic time.\par 

Another viewpoint on halo concentrations can be obtained by looking at the inferred probability distribution function (PDF) for $c(M,z)$. To this end, Fig. \ref{f_cPDF} juxtaposes normalised histograms of CDM concentrations vs. cFDM ones, again at redshift $z=4.38$. Following \cite{Jing_2002}, we fit the PDFs with a phenomenological lognormal distribution
\begin{equation}
p(c)\mathrm{d}c = \frac{1}{\sqrt{2\pi}\sigma_{c}}\exp\left[-\frac{(\ln c - \ln \bar{c})^2}{2\sigma_c^2}\right] \mathrm{d}\ln c,
\end{equation}
which we find to perform well across all mass bins, cFDM particle masses and again for redshifts in the range $3.4 \leq z \leq 6.3$. The scatter in $c(M,z)$ around the median relations has been investigated by other authors. While \cite{Diemer_2015} reports a CDM scatter of $0.16$ dex at all redshifts and masses using a halo inventory that includes unrelaxed halos, \cite{Neto_2007} finds $\sigma_{\log_{10}c}\sim 0.1$ in a selection of mass bins for relaxed halos at $z=0$. At $z=4.38$, we identify a CDM scatter of $\sigma_c \sim 1.32$ in the highest mass bin, which translates to $\sigma_{\log_{10}c}\sim 0.12$ and is thus consistent. The monotonicity of concentrations in CDM is evident from the PDFs, as in the analogous, bottom panel of Fig. \ref{f_cVersusM}.\par 

We take a moment to note that the agreement we find with \cite{Ludlow_2016} is conditional on adopting (most of) their selection functions, their choices for calculating halo density profiles and their Einasto fitting procedure. Most importantly, by discarding unrelaxed halos the median $c(M,z)$ relation in CDM becomes strictly monotonically decreasing with halo mass. If we were to include unrelaxed halos, we would encounter an ``upturn'' in the concentration of high-mass halos. Rather than `LB 16, CDM', a much better fitting formula would then be provided by \cite{Diemer_2019}. While their analytic framework naturally incorporates unrelaxed halos in CDM, its extension to cFDM would necessitate a new parameter in the model. Since the upturn at the high-mass end is largely insensitive to a primordial power spectrum truncation, we remove the upturn and accept a bias toward dynamically older systems.\par

Another controversy concerns the very definition of halo concentration for a three-parameter model such as the Einasto profile. For the latter, the parameter $\alpha$ co-determines not only the shape of the profile but also the density of the halo at its centre. An intuitive definition of halo concentration $c$ mapping more dense centres to higher concentrations must make $c$ a function of both $R_{200}/r_{-2}$ and $\alpha$. For instance, the ratio of maximum to virial circular velocities can be cast into a concentration by assuming a profile parametrisation as is done in \cite{Klypin_2016}, but it can also be used as an alternative method for characterising the halo concentration that does not assume any density profile \citep{Gao_2004}. Other refinements could be added by considering ellipsoidal profiles rather than spherical profiles (cf. Section \ref{s_data_availability}), in which case we would take into account the shape of the dark matter particle distribution, constraining the bins to follow the main directions of this distribution. When using ellipsoidal profiles, fitted masses and concentrations thus tend to be higher \citep{Gonzalez_2022}.

Since our goal is to generalise and extend density profile results found for WDM to cFDM, we choose the simple $c=R_{200}/r_{-2}$ definition, contend with spherically averaged profiles, and exclude unrelaxed halos, all in agreement with \cite{Ludlow_2016}.

\subsection{Shape Statistics}
\label{ss_shape_stats}

The lack of a fully self-consistent theory of DM halo structure formation let alone galaxy formation motivates the use of cosmological simulations. This is even more true for alternative DM scenarios, for which many seminumerical methods such as a generalised halo model have only recently been introduced \citep{Schneider_2012}.\par 

It is known that halo growth is anisotropic by virtue of accretion being clumpy and directional, such as along filaments and sheets. While the anisotropy is known to result in non-spherical halos \citep{Maccio_2008}, statistical properties of cFDM halo shapes have not yet been investigated, to the best of our knowledge. Here, we fill this gap by focusing on how halo shapes are affected by gradually decreasing the cFDM particle mass $m$.\par 

Shape profiles out to $5 R_{\text{vir}}$ are calculated by applying the mode-centred Katz-Dubinski algorithm of Section \ref{ss_shape_finding} on the central subhalos.  Fig. \ref{f_shapes} compares CDM and cFDM median shape profiles for $N$-body, $1024^3$ resolution, $L_{\text{box}} = 40$ cMpc$/h$ runs at redshift $z=4.38$. We highlight beyond-virial radius regions in the plot as they ought to be interpreted as part of the cosmic environment rather than the halos per se. Let us begin by outlining some shape characteristics of CDM halos. We find that halos are least spherical near the halo centre, with axis ratios $\langle q \rangle \sim 0.65$ and $\langle s \rangle \sim 0.45$ for intermediate-mass halos at $R_{15} = 0.15 \ R_{\text{vir}}$. As a comparison, the same group of halos is significantly more spherical near the virial radius $R_{\text{vir}}$ with $\langle q \rangle \sim 0.70$ and $\langle s \rangle \sim 0.50$.\par

We speculate that the monotonicity of $q$ and $s$ as a function of ellipsoidal radius is a property of hierarchical structure formation, i.e. occurs in cosmological models in which the dimensionless matter power spectrum $\Delta^2(k)=k^3P(k)/(2\pi^2)$ increases with wavenumber $k$. In theory, it should thus be predictable using (semi\=/)analytical arguments. Similar to halo density profiles, hierarchical clustering clearly does not produce objects with a universal shape, but the \textit{distributions} may be close to universal\footnote{Strictly speaking, the monotonicity can only be considered a universality if the dependence on at least the redshift and cosmological parameters of the $q$- and $s$-profiles can be subsumed via an appropriate rescaling. Since we do not investigate this claim thoroughly, we will henceforth refrain from calling the monotonicity a universality.}. Spherically averaged halo density profiles $\rho(r)$ are considered universal if they are purely a function of total halo mass $M$. Current efforts to understand the $c$-$M$ relation are precisely tackling the question of universality since the latter is broken unless concentration can itself be described as a function of mass. \cite{Klypin_2016} find that in the so-called \textit{plateau regime}, concentration does not depend on halo mass. A potential universality for shapes is underinvestigated as of yet, most likely due to its higher level of complexity arising from 3D considerations as opposed to 1D. We advocate for more emphasis thereon as it is beyond the scope of this work.\par

As with density profiles, special care is needed to understand how non-relaxed halos modify any such monotonicity. For our shape analysis, including Fig. \ref{f_shapes}, we do not impose any relaxedness condition, which might be the reason why we find the monotonicity in $q$ and $s$ to break for the highest-mass halos, which are expected to undergo many mergers and are still in the \textit{fast-accretion regime} in the nomenclature of \cite{Diemer_2019}. On a similar note, how does monotonicity in $q$ and $s$ change in the low-redshift Universe of $z\lesssim 4$? As $N$-body simulations including our own ones indicate, the monotonicity becomes a steeper one as time evolves\footnote{If we were to include baryons and radiative feedback processes, the steepening would be strongly impeded and the monotonicity eventually broken at lower redshifts \citep{Chua_2019, Chua_2021}, see also Section \ref{ss_shapes_halos_gxs}.}.\par 

Another consequence of the higher rate of mergers of high-mass CDM halos that we just hinted at is the enhanced prolateness of such halos, already identified by \cite{Allgood_2006}. At $z=4.38$, low-mass halos with $M = 10^{9}-10^{10} M_{\odot}/h$ have a triaxiality of $\langle T \rangle \sim 0.70$ at the virial radius while high-mass halos are more elongated with $\langle T \rangle \sim 0.80$.

\subsubsection{Impact of Cosmic Filaments on cFDM Halos} 
\begin{figure*}
\hspace{-0.8 cm}
\includegraphics[scale=0.67]{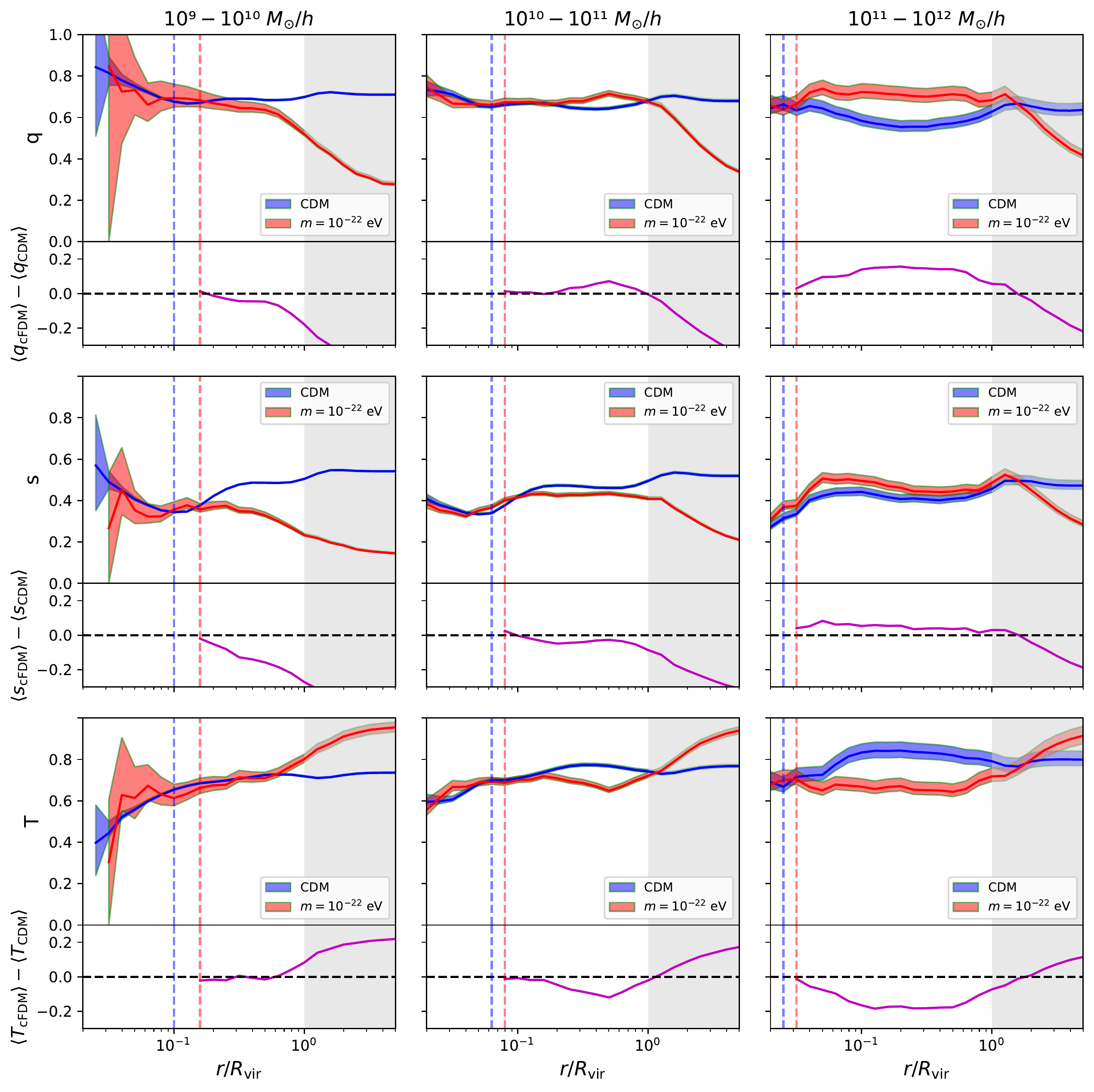}
\caption{Shape profiles in different cosmologies, for $N$-body, $1024^3$ resolution, $L_{\text{box}} = 40$ cMpc$/h$ runs at redshift $z=4.38$. We show the intermediate-to-major axis ratio $q$ (first row), the minor-to-major axis ratio $s$ (second row) and the triaxiality $T$ (third row) profiles for CDM (blue) as compared to cFDM (red) with particle mass $m=10^{-22}$ eV for halo mass bins $10^{9}-10^{10} M_{\odot}/h$ (first column), $10^{10}-10^{11} M_{\odot}/h$ (second column) and $10^{11}-10^{12} M_{\odot}/h$ (third column). Median difference curves are drawn in solid purple. Halo-to-halo variations are represented by shaded regions which enclose the $25$th-$75$th percentiles. Dashed blue and red lines delineate $\kappa = 7$ lower convergence radii. The shaded light-grey region denotes the ambient cosmic environment of the halos.}
\label{f_shapes}
\end{figure*}

How does a power spectrum cutoff affect halo shapes? As we see in Fig. \ref{f_shapes} for the rather extreme $m=10^{-22}$ eV model, cFDM halos are less oblate than their CDM counterparts beyond and around the virial radius $R_{\text{vir}}$. The lower values of $q$ and $s$ are indicative of the increased role cosmic filaments play in cFDM, into which most halos at $z=4.38$ are embedded. As we will show in future work, cosmic filaments in cFDM feature higher mean overdensities than in CDM and in relative terms contribute more to the build-up of large-scale tidal forces. This leads to an elongation (higher $T$-values) of DM halos at large ellipsoidal radii. Their profound gravitational influence renders small-mass halos less spherical (reduced $q$ and $s$) even down to deeper regions around $R_{15}$. High-mass cFDM halos of mass $M = 10^{11}-10^{12} M_{\odot}/h$ are less disrupted by cosmic filaments as their deeper gravitational potential wells provide greater shielding.\par

\begin{figure*}
\hspace{-0.4cm}
\includegraphics[scale=0.7]{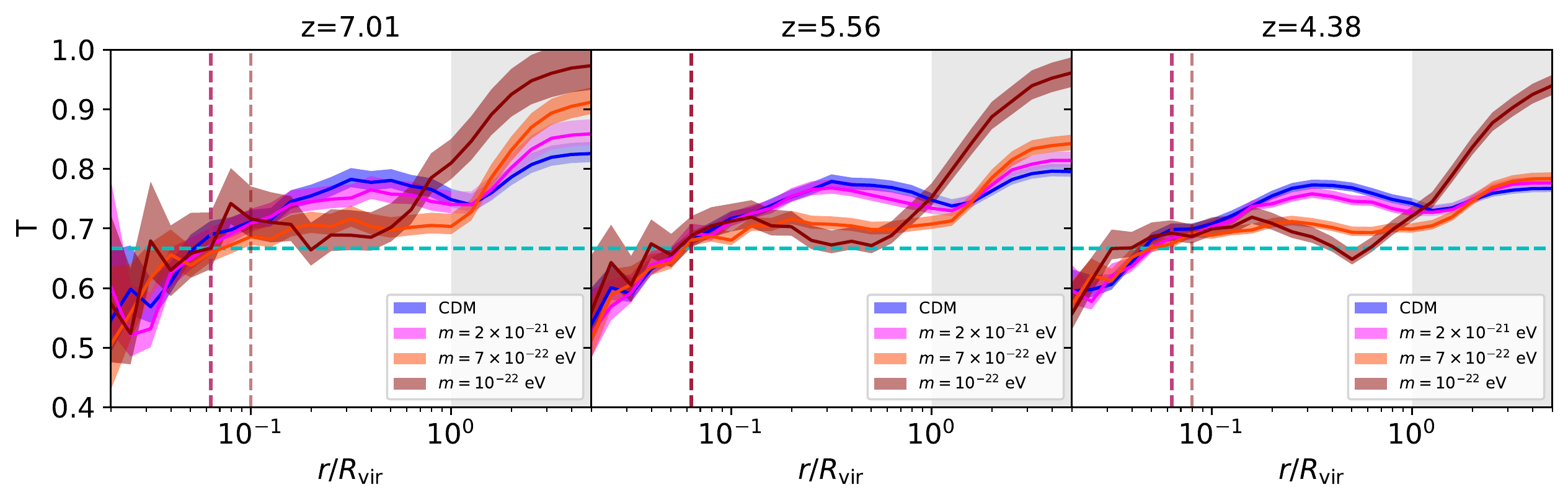}
\caption{Triaxiality profiles in different cosmologies, for $N$-body, $1024^3$ resolution, $L_{\text{box}} = 40$ cMpc$/h$ runs. We compare CDM (blue) to cFDM with particle mass $m=2\times 10^{-21}$ eV (magenta), $m=7\times 10^{-22}$ eV (dark-red) and $m=10^{-22}$ eV (orange-red) across redshifts indicated above the columns. Shaded regions enclose the $25$th-$75$th percentiles. The results are for intermediate-mass halos in the range $10^{10}-10^{11} M_{\odot}/h$. Vertical dashed lines delineate $\kappa = 7$ lower convergence radii. The cyan-colored dashed horizontal line at $T=2/3$ denotes the triaxial-prolate transition.}
\label{f_t_shapes}
\end{figure*}

To appreciate the redshift- and cFDM particle mass-dependence of shape profiles, Fig. \ref{f_t_shapes} shows triaxiality curves over a wider range in parameter space. We find that the influence of cosmic filaments on shape profiles fades away as time evolves. Beyond-virial radius profiles of cFDM halos for $m=2\times 10^{-21}$ eV closely match their CDM counterparts around $z=5.56$, while the profiles do not show much resemblance even at $z=4.38$ for the more extreme $m=10^{-22}$ eV run. This suggests that with power redistributing among the various scales due to non-linear structure formation, cFDM / FDM filaments eventually fragment over time into less prolate structures \citep{Smith_2011, Mocz_2019}. This complicated process takes place over a redshift range that is hard to pinpoint, as the fragmentation time of a filament depends on its core radius $r_0$, amongst others. The latter is a good proxy for where the simple model of isothermal cylinders in hydrostatic equilibrium breaks down \citep{Ramsoy_2021}. Some filaments prove stable enough so as to not break up at all. Overall though, the effect of the power spectrum cutoff becomes less noticeable at lower redshifts. Since structure formation is predominantly linear and thus more pristine at higher redshift, we conclude that with increasing redshift it is easier to distinguish cFDM from CDM. This is insofar as the theoretically relevant but not directly observable halo shapes are concerned. An accurate high-$z$ determination of galaxy shapes and a thorough understanding of their dependence on parent halo shapes thus might shed light on the nature of DM.\par

Enhanced prolateness of halos is expected to be observed in WDM simulations as well, since thermal velocities are incapable of modifying halo shapes (in addition to the linear effects stemming from the matter power spectrum suppression) beyond a very small core with size of a few parsecs for $m_X\sim 1$ keV \citep{Maccio_2012}. Hence, we find that a combination of anisotropic gravity and a primordial power spectrum cutoff results in an elongation of the cosmic web at high-$z$, which is most noticeable for halos and filaments. Crudely speaking, the cutoff-mediated elongation of filaments is passed on to the embedded halos via tidal shearing, until the tenuous filaments eventually break up around $z\sim 6$ for $m\sim 7\times 10^{-22}$ eV and elongation fades away with non-linear structure formation, cf. \cite{Mocz_2019}.

\subsubsection{Anticorrelation between Sphericity and Halo Mass} 
We confirm the positive correlation between triaxiality $T$ and halo mass $M$ at the virial radius $R_{\text{vir}}$. It is more customary though to quantify the correlation of the median sphericity $\langle s \rangle$ with halo mass $M$. Similar to the results of \cite{Chua_2019, Butsky_2016}, we find a negative correlation of $\langle s \rangle$ with $M$ for CDM halos. Following \cite{Allgood_2006}, we express the sphericitiy-mass relation as
\begin{equation}
\langle s \rangle = a\left(\frac{M_{\text{vir}}}{M_{\ast}(z)}\right)^b,
\end{equation}
where $M_{\text{vir}}$ represents the virial mass and $M_{\ast}(z)$ is the cosmology-dependent characteristic non-linear mass for redshift $z$ such that the RMS real-space top-hat smoothed overdensity at scale $R(M_{\ast}) = (3M_{\ast}/(4\pi \rho_{\text{crit}} \Omega_m))^{1/3}$ is $\delta_c = 1.68$. In other words,
\begin{equation}
\sigma^2(M_{\ast}(z), z) = \int_0^{\infty}\frac{dk}{k}\left[\frac{3j_1(kR(M_{\ast}))}{kR(M_{\ast})}\right]^2\Delta^2(k,z) \stackrel{!}{=} \delta_c^2,
\label{e_mstar_def}
\end{equation}
where $j_1(x)$ is a spherical Bessel function of the first kind and $\Delta^2(k,z) = k^3P^{\text{lin}}(k,z)/(2\pi^2)$. \cite{Allgood_2006} found that the dependence of the distributions of halo shapes on the amplitude of density perturbations, $\sigma_8$, was well described by the cosmology dependence of $M_{\ast}$ alone. We remind the reader that the power spectrum $P^{\text{lin}}(k,z)$ that appears in the definition of the variance $\sigma^2$ refers to the \textit{linear} power spectrum in the corresponding cosmology, extrapolated to possibly low redshift $z$. The same remark holds true for the definition of $\sigma_{8} \coloneqq \left[\sigma^2(8\ \text{Mpc}/h, z=0)\right]^{\frac{1}{2}}$. In $\Lambda$CDM, $M_{\ast}(z=0) = 4.66\times 10^{12}M_{\odot}/h$, $M_{\ast}(z=1) = 1.54 \times 10^{11}M_{\odot}/h$ and $M_{\ast}(z=3.41) = 8.84 \times 10^{7}M_{\odot}/h$.\par 

For cFDM cosmologies, one could resort to the \cite{Viel_2005} parametrisation of the linear WDM-to-CDM transfer function ratio, since scales beyond $k=10$ $h$/Mpc are largely suppressed in the integrand of Eq. \eqref{e_mstar_def} by the Bessel factor, while the linear cFDM power spectrum below $k=10$ $h$/Mpc can be well approximated by the \cite{Viel_2005} parametrisation. However, for pure cFDM and WDM cosmologies $\mathcal{C}$, $M_{\ast}(z, \mathcal{C})$ ceases to exist for larger redshifts, e.g. $z \gtrsim 2$ for $m = 10^{-22}$ eV or equivalently $m_{\text{WDM}} = 0.84$ keV. The reason is that $\sigma^2(M, z)$ flattens off for small $M$, instead of diverging as $\lim_{M \to 0} \sigma^2(M, z) \rightarrow \infty$ as happens for cosmologies with a high-$k$ power index $n \geqslant -3$ such as $\Lambda$CDM.\par 

\renewcommand{\arraystretch}{2}
\begin{table*}
	\centering
    \begin{tabular}{c c c | c c | c c | c c} 
     & \multicolumn{2}{c|}{CDM} & \multicolumn{2}{c|}{$m=2\times 10^{-21}$ eV} & \multicolumn{2}{c|}{$m=7\times 10^{-22}$ eV} & \multicolumn{2}{c}{$m=10^{-22}$ eV} \\ \cline{2-9}
     & a & b & a & b & a & b & a & b \\ \hline \hline
    \multicolumn{1}{c|}{$q_{vir}$} & 0.759 & -0.012 & 0.768 & -0.009 & 0.740 & -0.004 & 0.437 & 0.069 \\ \hline
    \multicolumn{1}{c|}{$q_{15}$} & 0.648 & 0.006 & 0.652 & 0.005 & 0.631 &  0.012 & 0.494 & 0.044 \\ \hline
    \multicolumn{1}{c|}{$s_{vir}$} & 0.541 & -0.011 & 0.521 & -0.006 & 0.511 & -0.005 & 0.201 & 0.124 \\
\hline
    \multicolumn{1}{c|}{$s_{15}$} & 0.289 & 0.081 & 0.276 & 0.080 & 0.267 & 0.089 & 0.233 & 0.098
    \end{tabular}
    \caption{Fitting parameters to the mass-dependency equation $\langle p \rangle = a(M_{\text{vir}}/M_{\ast}(z=3.41))^b$ for two different radii $R_{\text{vir}}$ and $R_{15}$, with $p=q$ or $s$. Results are shown for $N$-body, $1024^3$ resolution, $L_{\text{box}} = 40$ cMpc$/h$ runs.}
    \label{t_ab_fits}
\end{table*}
\renewcommand{\arraystretch}{1}

In addition, $M_{\ast}(z, \mathcal{C} = \Lambda\text{CDM})$ drops quickly below the resolution limit $m_{\text{DM}} = 5.11 \times 10^6 M_{\odot}/h$ of the $N=1024^3$, $L_{\text{box}} = 40$ cMpc$/h$ runs at around $z \gtrsim 4$. Table \ref{t_ab_fits} thus shows best-fit parameters for $z=3.41$ only, with $M_{\ast}(z, \mathcal{C} = \Lambda\text{CDM})$ values adopted for the cFDM cosmologies as well. \par

We confirm the anticorrelation ($b < 0$) between sphericity and halo mass that has been found previously by other authors \citep{Chua_2019, Allgood_2006} and extend the result to moderate values of the cFDM particle mass $m$. What is the physical origin of this phenomenon? Again, we expect halos with masses above $M_{\ast}$ to undergo a higher rate of mergers than halos with masses below $M_{\ast}$. Since it has been shown that this merging happens along preferred directions \citep{Zentner_2005}, the enhanced prolateness and reduced sphericity is primarily due to merging. Halo merging thus plays a significant role in the distribution of shapes.\par 

For particle mass $m=10^{-22}$ eV, however, we observe that the anticorrelation between sphericity and halo mass breaks down and turns into a positive correlation. This is true for both $b=0.069$ at $q_{\text{vir}}$ and $b=0.124$ at $s_{vir}$. We find no clear anti-correlation for CDM in the central regions around $R_{15}$ as opposed to some low-redshift results, e.g. \cite{Allgood_2006, Chua_2019}. However, the magnitude of the power index $b$ is still preferentially increased by the power spectrum cutoff, up to $b=0.098$ at $s_{15}$ for cFDM with $m=10^{-22}$ eV. Some authors have reported no significant change in the magnitude of the power index $b$ across redshift, with some even claiming a steepening with redshift \citep{Jing_2002}. However, Fig. 2 in \cite{Allgood_2006} hints at the opposite trend, which we can confirm. In fact, the magnitude of $b$ is so low at $z\gtrsim 4$ that the fitting exponent is consistent with zero in CDM, another reason against pursuing such a fitting procedure at too high redshifts. 

\subsubsection{PDF of Halo Shapes} 
\begin{figure*}
\includegraphics[scale=0.65]{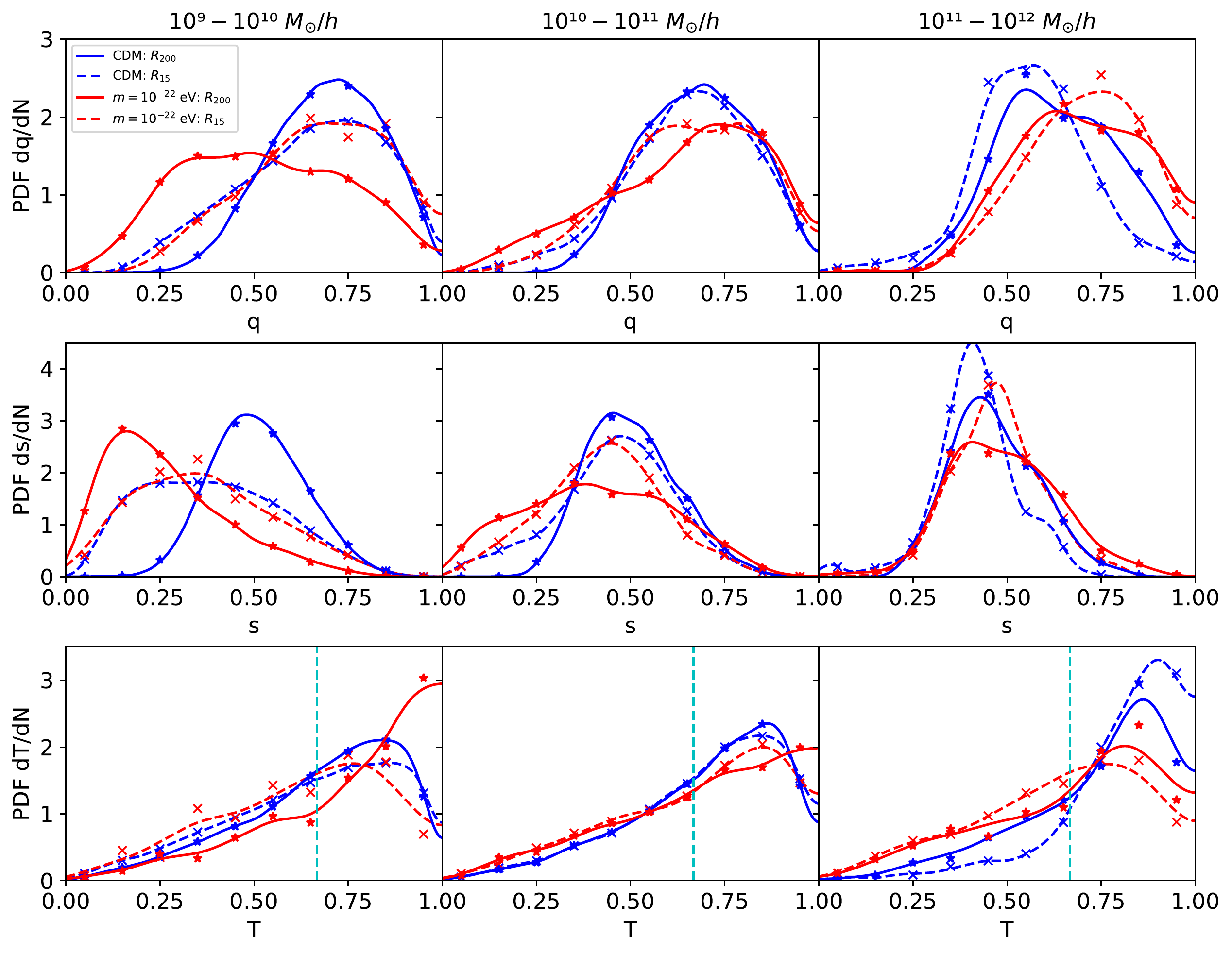}
\caption{Probability density functions of the shape parameters in different cosmologies, for $N$-body, $1024^3$ resolution, $L_{\text{box}} = 40$ cMpc$/h$ runs, at redshift $z=4.38$. We compare the intermediate-to-major axis ratio $q$ (top row), the minor-to-major axis ratio $s$ (middle row) and the triaxiality $T$ (bottom row) PDFs in cFDM with particle mass $m=10^{-22}$ eV (red) to CDM (blue) at $R_{200}$ (solid lines) and $R_{15}$ (dashed lines). Mass bins are $10^{9}-10^{10} M_{\odot}/h$ (left column), $10^{10}-10^{11} M_{\odot}/h$ (middle column) and $10^{11}-10^{12} M_{\odot}/h$ (right column). The cyan-colored dashed vertical line at $T=2/3$ characterises the triaxial-prolate transition. While the symbols denote normalised histogram values, the curves plot Gaussian kernel density estimates thereof.}
\label{f_qsT}
\end{figure*}

For the PDFs of the axis ratios, we recover the CDM trends already identified in \cite{Schneider_2012_2} at lower redshifts: As Fig. \ref{f_qsT} indicates for $z=4.38$, sphericity $s$ follows an approximately symmetrical distribution, with cFDM halos featuring lower values especially beyond and around $R_{\text{vir}}$. Lowest-mass halos in cFDM, however, peak at sphericities as low as $s\sim 0.15$, indicating that the majority of these halos is highly non-spherical. We note that $T$ is negatively skewed, suggesting that there is a heavy tail of oblate and triaxial halos in both CDM and cFDM scenarios. As in Fig. \ref{f_t_shapes}, we find a reduced prolateness (smaller median of $T$) of intermediate- and especially high-mass cFDM halos compared to the CDM ones around $R_{15}$. We see that for intermediate-mass cFDM halos, this can be traced to an increased median in the $q$-distribution and a reduced median in the $s$-distribution. For high-mass cFDM halos, it primarily stems from an increased median in the $q$-distribution.\par 

The nature of DM thus has strong imprints on halo shapes, most pronounced for the smallest explored halos at $R_{15}$. Specifically, such halos in cFDM are less spherical (smaller $s$ and $q$) while having an overall higher triaxiality (larger $T$). It is possible that given a primordial power spectrum and expansion history of the Universe, one could estimate the shape distribution of virialised objects at a given redshift $z$ (semi\=/)analytically and thus theorise for instance the smaller median of $T$ for intermediate- and high-mass halos in cFDM cosmologies. To the best of our knowledge, such an analysis has yet to be performed.

\subsection{Intrinsic Alignment}
\label{ss_ia}

The reason why intrinsic alignments have gained considerable attention in recent years is twofold. First, the next generation of galaxy weak lensing surveys such as the \href{http://www.euclid-ec.org/}{\scshape{Euclid}\normalfont} mission will suffer from highly biased cosmological inferences without a proper alignment modelling informed by simulations and seminumerical methods. At the same time, intrinsic alignments of galaxies provide complementary cosmological information on their own, assuming they can be disentangled from the lensing signal by means of either galaxy color \citep{Yao_2020} or polarisation data \citep{Brown_2011}.\par 

To remain consistent throughout this work, we focus on the linear intrinsic alignment strengths of DM halos and the impact of a power spectrum cutoff thereon. One should be careful when carrying over our conclusions for the linear alignment strengths of halos to the population of elliptical galaxies they may host. Ellipticals have more spherical gas distributions than the underlying DM halo, which is a corollary of the \textit{X-ray shape theorem} \citep{Buote_1994}, since the gravitational potential represents an overall average of the local density profile. Gas pressure is also more isotropic than the anisotropic velocity ellipsoids of a DM halo, further contributing to more spherical gas distributions. These theoretical arguments are corroborated by various observations \citep{Buote_1998, Morandi_2010}. Moreover, the misalignment between ellipticals and their DM halos can pose additional problems in case such a misalignment is correlated with the local tidal fields rather than being purely random. To make matters worse, the linear intrinsic alignment strength of spiral galaxies is expected to be about an order of magnitude smaller than those for ellipticals \citep{Zjupa_2020}.

\subsubsection{Geometric Alignments} 
The simplest way to quantify the intrinsic alignment of halos is via geometry. One measure is shape-position alignment, which we have introduced in Section \ref{ss_def_align}. It can be considered a special case of the more general notion of ellipticity-direction cross-correlation \citep{Lee_2008}. Likewise, the shape-shape alignment measure can be mapped onto the ellipticity-ellipticity correlation function.\par

\begin{figure}
\hspace{-0.5cm}
\includegraphics[scale=0.5]{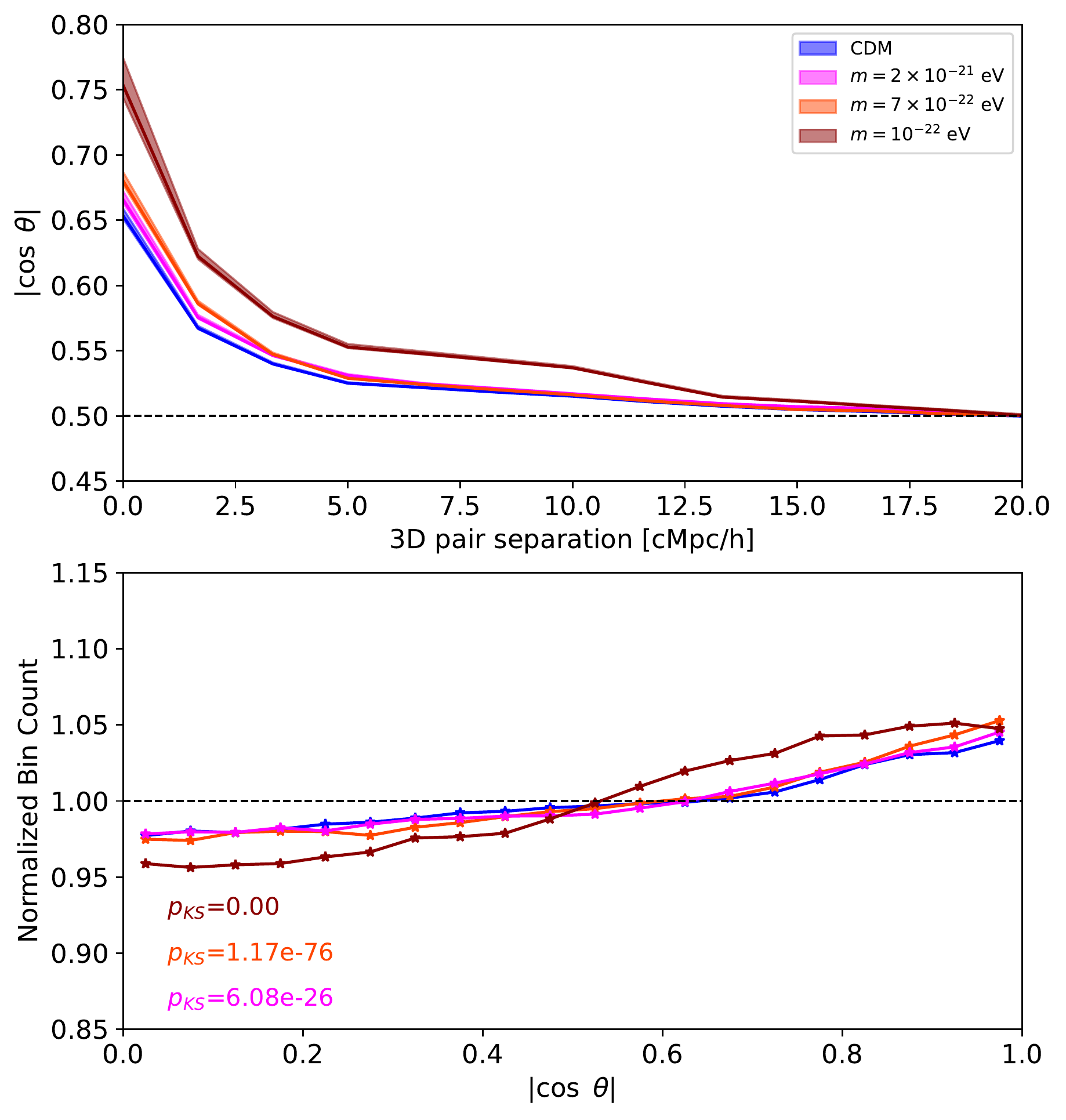}
\caption{High-$z$ shape-position alignment statistics in different cosmologies, for $N$-body, $1024^3$ resolution, $L_{\text{box}} = 40$ cMpc$/h$ runs, at redshift $z=4.38$. We compare CDM halos (blue) to cFDM halos with particle mass $m=2\times 10^{-21}$ eV (magenta), $m=7\times 10^{-22}$ eV (dark-red) and $m=10^{-22}$ eV (orange-red). The results are for intermediate-mass halos in the range $10^{10}-10^{11} M_{\odot}/h$. Top: Shape-position alignments vs. 3D pair separation, out to $L_{\text{box}}/2=20$ cMpc$/h$ to avoid geometric bias induced by the simulation box. Shaded regions enclose the $25$th-$75$th percentiles. Bottom: PDF of shape-position alignments, with results of the Kolmogorov-Smirnov test of independence between the CDM and cFDM samples.}
\label{f_alignments}
\end{figure}

In Fig. \ref{f_alignments} we present shape-position alignment statistics for CDM as compared to cFDM halos at redshift $z=4.38$, restricted to mass bin $10^{10}-10^{11} M_{\odot}/h$. We find that for large pair separations, the median value for $|\cos \theta|$ asymptotically approaches $0.5$, a value consistent with a uniform random shape orientation and halo clustering. For small pair separations, intermediate-mass CDM halos conspire to a median $|\cos \theta|$ value of around $0.653^{+0.006}_{-0.002}$, which is in very close agreement with the value \cite{Pandya_2019} found for real-space 3D mock light cones that they generated for the CANDELS survey at lower redshifts of $z=1.0-2.5$. What we can also confirm (not shown) is that the shape-shape alignment strength barely reaches values of $|\cos \theta|\sim 0.6$ for smallest-separation halo pairs, which is systematically lower than the shape-position alignment magnitude.\par 

For cFDM cosmologies, Fig. \ref{f_alignments} clearly demonstrates how the geometric alignment is gradually enhanced as the cFDM particle mass $m$ is decreased. For our most extreme cFDM model with $m=10^{-22}$ eV, the median $|\cos \theta|$ curve climbs up to $0.753^{+0.021}_{-0.010}$, the wider-spread percentiles effected by the smaller number of halos compared to CDM. The enhancement is due to the lack of small-scale structure such as smaller-mass halos that would regularise the tidal fields generated by quasi-linear cosmic filaments and to a lesser extent 2D cosmic sheets. Shape-position alignments are significantly different from the uniform random result even out to 3D pair separations of $10$ cMpc/$h$, where $|\cos \theta| \sim 0.54$. Upon inspection of the $|\cos \theta|$ distribution after marginalising over 3D pair separations, we find that the Kolmogorov-Smirnov $p$-value, $p_{\text{KS}}$, for rejecting the hypothesis that the cFDM and CDM signals are drawn from the same distribution is very low. In other words, the hypothesis can be ruled out with very high significance. As expected, $p_{\text{KS}}$ increases with $m$.

\subsubsection{Intrinsic Alignment Modelling}
While the linear alignment model (LAM) introduced in \cite{Hirata_2004} and \cite{Catelan_2001} is traditionally applied to elliptical galaxies, the condition of a virialised, velocity dispersion-stabilised system is also satisfied by halos. To keep our focus on the properties of the cosmic web and to harness greater statistical power, we investigate intrinsic alignment correlations of halos within the LAM. In this model, the halo is assumed to be spherically symmetric in isolation, perturbed by the presence of a cosmic tidal field. Let $\sigma^2(r)$ denote the isotropic velocity dispersion of DM resolution elements in the unperturbed halo with $r=|\mathbf{r}| = |\mathbf{x}-\mathbf{x}_0|$ being the distance of the particle to the centre of mass of the halo. In a steady-state Jeans equilibrium one has
\begin{equation}
\frac{1}{\rho} \frac{\D (\sigma^2\rho)}{\D r} = - \frac{\D \Phi_{\text{eq}}}{\D r},
\end{equation}
with denstiy $\rho(r)$ and gravitational potential $\Phi_{\text{eq}}(r)$. The Jeans equilibrium density profile is thus
\begin{equation}
\rho(r) = \rho_0 \cdot \exp\left(-\frac{\Phi_{\text{eq}}(r)}{\sigma^2}\right),
\label{e_eq_rho_prof}
\end{equation}
with a constant $\rho_0$ if $\sigma^2$ is constant with radius.\par 

In the presence of an anisotropic gravitational tidal field surrounding the halo, the halo will find a new dynamical equilibrium within the free fall time scale $\varpropto 1/\sqrt{G\rho}$. As the halo is virialised, setting $\rho = 200 \Omega_m\rho_{\mathrm{crit}}$ with $\rho_{\mathrm{crit}} = 3H_0^2/(8\pi G)$ yields the fraction $\sim \sqrt{4\pi/300\Omega_m}$ of the Hubble time $1/H_0$ as the upper limit for the free fall time scale. Taylor-expanding the anisotropic large-scale gravitational potential $\Phi(\mathbf{r})$ to second order around the centre of mass of the halo gives
\begin{equation}
\Phi(\mathbf{r}) = \Phi(\mathbf{x}_0) + \Phi_{,a}(\mathbf{x}_0)r_a + \frac{1}{2}\Phi_{,ab}(\mathbf{x}_0)r_ar_b + ...,
\label{e_phi_taylor}
\end{equation}
where we use the Einstein summation and $_{,a}$ denotes derivation with respect to the position coordinate $r_a$. While the first derivative in Eq. \eqref{e_phi_taylor} accelerates the halo, providing it with a non-zero peculiar velocity, shape modifications are effected by higher order derivatives \citep{Ghosh_2021, Piras_2018, Giesel_2022}. Provided $\sigma^2$ remains unscathed by the perturbation and the halo is small compared to the curvature scale $\mathcal{S}$,
\begin{equation}
\frac{1}{\mathcal{S}^2} \coloneqq \bigg\lvert \frac{1}{\sigma^2}\frac{\D ^2}{\D r^2} \Phi(r)\big\rvert_{\mathbf{x}_0} \bigg\rvert,
\end{equation}
one can perturb the steady-state density profile $\rho(r)$ of Eq. \eqref{e_eq_rho_prof} to lowest order to obtain
\begin{equation}
\tilde{\rho}(\mathbf{r}) \sim \rho(r)\left(1-\frac{\Phi_{,ab}(\mathbf{x}_0)r_ar_b}{2\sigma^2}\right).
\end{equation}
The first three even moments of the density distribution are given by
\begin{align}
S_0 &\coloneqq \int \D^3 r \rho(\mathbf{r}), \\
\label{e_second_moment}
q_{ij} &\coloneqq \frac{1}{S_0} \int \D^3 \rho(\mathbf{r})r_ir_j,\\
s_{ijkl} &\coloneqq \frac{1}{S_0} \int \D^3 \rho(\mathbf{r})r_ir_jr_kr_l.
\end{align}
Eq. \eqref{e_second_moment} is the generalisation of Eq. \eqref{e_shape_tensor}  for the continuous case. The perturbation of the second moment $q_{ij}$ around $\mathbf{r}_0$ due to cosmic tides is thus
\begin{equation}
\Delta q_{ij}(\mathbf{x}_0) \coloneqq \tilde{q}_{ij}(\mathbf{x}_0)-q_{ij} = -\frac{1}{2\sigma^2}\Phi_{,ab}(\mathbf{x}_0)s_{ijkl}.
\label{e_q_perturb}
\end{equation}

The (in the case of halos) unobservable complex ellipticity in projection along the Cartesian $z$-axis follows from $\tilde{q}_{ij} \equiv \tilde{q}_{ij}(\mathbf{x}_0)$ as
\begin{equation}
\tilde{\epsilon} = \frac{\tilde{q}_{xx}-\tilde{q}_{yy}}{\tilde{q}}+i\frac{2\tilde{q}_{xy}}{\tilde{q}},
\label{e_eps_tilde}
\end{equation}
where the halo size $\tilde{q} = \tilde{q}_{xx}+\tilde{q}_{yy} = q$ is assumed to be constant. By inserting the tidal shape perturbation \eqref{e_q_perturb} we obtain
\begin{equation}
\tilde{\epsilon} = \epsilon_0 - \frac{1}{2\sigma^2q}\sum_{i,j \in \lbrace x,y\rbrace}\Phi_{,ij}(s_{ijxx} - s_{ijyy} + 2is_{ijxy}).
\end{equation}
We note here that by symmetry (`matching the indices'), Eq. \eqref{e_q_perturb} must hold to lowest order even if we do not assume an unperturbed halo that is spherically symmetric, $q_{ij} = \frac{q}{2}\delta_{ij}$. For now we generalize the unperturbed halo to being perturbatively isotropic, i.e. $q_{ij} = \frac{q}{2}\delta_{ij} + \eta_{ij}$, in which case the leading order contribution becomes
\begin{equation}
\tilde{\epsilon} = \epsilon_0 - D(\Phi_{,xx} - \Phi_{,yy} + 2i\Phi_{,xy}) \eqqcolon \epsilon_0 - D(T_{+} + iT_{\times}).
\end{equation}
The constant
\begin{equation}
D \coloneqq \frac{c/q+q/4}{\sigma^2}
\end{equation}
absorbs the halo properties, i.e. the concentration or peakiness $\tilde{c} = c$ which is the fourth cumulant of the density $\rho(\mathbf{r})$, the velocity dispersion $\sigma^2$ and its size $q$.\par 

In a statistical sample such as halos from a simulation box, the intrinsic ellipticity $\epsilon$ can be modelled by a random variable with zero mean and some amount of dispersion, such that after dropping the tilde one obtains
\begin{equation}
\epsilon = \epsilon_{+} + i\epsilon_{\times} = -D(T_{+} + iT_{\times}),
\label{e_lam}
\end{equation}
which constitutes the essence of LAM. Eq. \eqref{e_lam} is reminiscent of gravitational lensing, and indeed the analogy is justified as intrinsic aligmnent is governed by the second derivatives of $\Phi/\sigma^2$ whereas lensing is governed by the second derivatives of $\Phi/c^2$. $D$ is strictly positive and is traditionally quoted in units of $c^2$, since it is beneficial to work with the dimensionless gravitational potential $\Phi/c^2$.\par 

To obtain $D$ for our simulation samples, we first compute the discrete version of the ellipticity \eqref{e_eps_tilde} for each halo. Next, we calculate the local gravitational tidal field 
\begin{equation}
\frac{\Phi_{,ij}}{c^2}(\mathbf{x}) = \frac{\partial^2\Phi(\mathbf{x})}{c^2\partial x_i\partial x_j}
\end{equation}
at the centre of mass of each halo. This can be achieved by first determining the three-dimensional overdensity field $\delta(\mathbf{x})$ for the full simulation volume using a CIC interpolation. We can then use a discrete Fourier transform to solve Poisson's equation and obtain the Hessian
of the potential algebraically in Fourier space via
\begin{equation}
\frac{\Phi_{,ij}}{c^2}(\mathbf{x}) = \frac{3\Omega_m}{2\chi_H^2a}\mathcal{F}^{-1}\left[ \frac{k_ik_j}{|\mathbf{k}|^2} \exp\left(-\frac{1}{2}|\mathbf{k}|^2\lambda^2\right)\mathcal{F}[\delta(\mathbf{x})]\right],
\end{equation}
where $\mathbf{k}$ is the comoving wave vector, $\chi_H \coloneqq \frac{c}{H_0}$ is the Hubble-distance and $a$ the cosmic scale factor. There is a non-removable uncertainty on the scale the effective tidal field relevant for LAM in Eq. \eqref{e_lam} should be evaluated on. Here, we consider a Gaussian smoothing scale of $\lambda =$ 1 cMpc/$h$. This scale corresponds to halos of mass $M = 4\pi/3 \Omega_m \rho_{\mathrm{crit}} \lambda^3 \sim 10^{11} M_{\odot}/h$. Finally, to obtain the tidal shear at the halo positions, we perform an inverse CIC interpolation, effectively increasing the smoothing scale $\lambda$ by about one grid cell.\par

\begin{figure*}
\hspace{-0.3cm}
\includegraphics[scale=0.49]{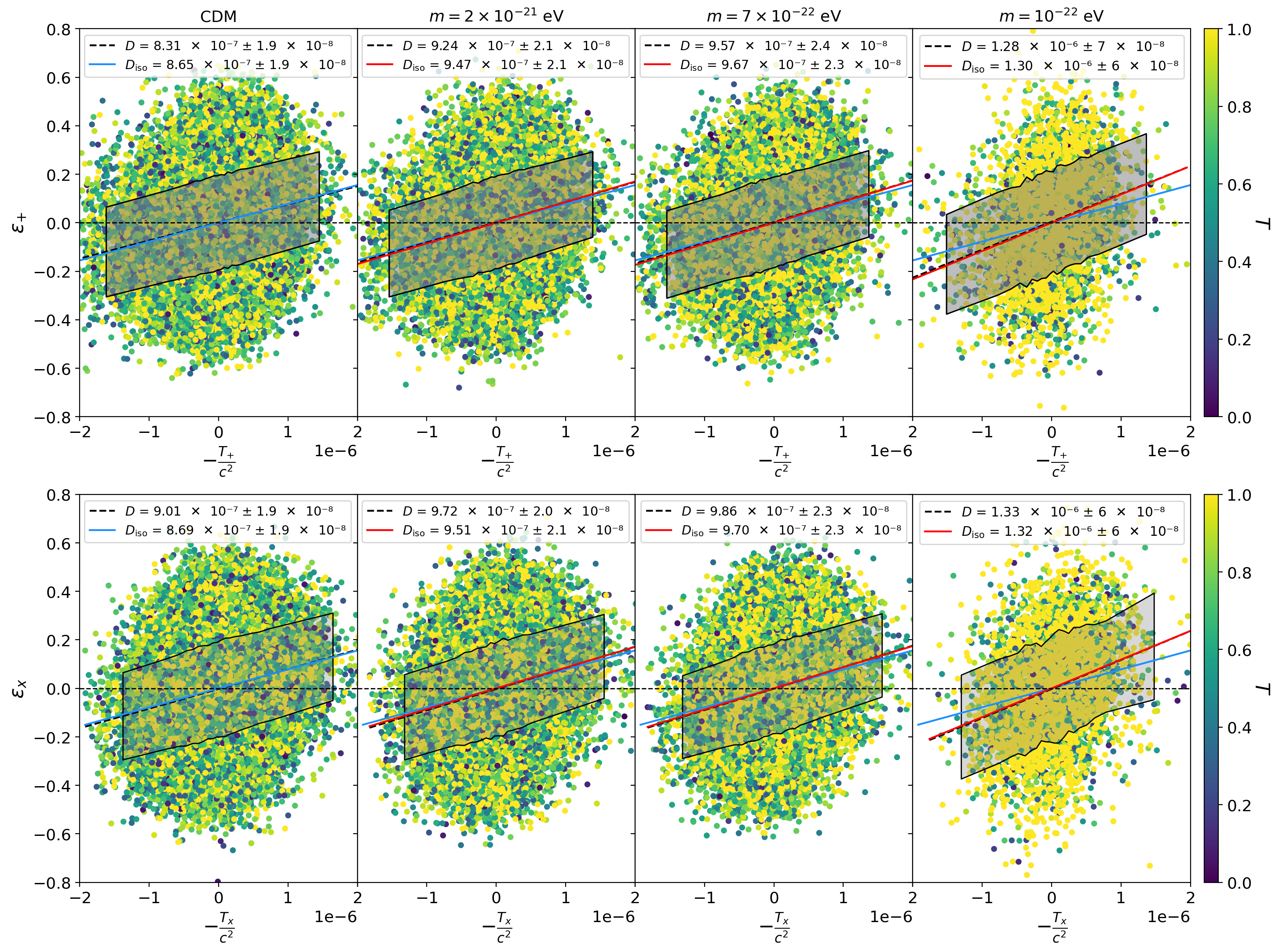}
\caption{Intrinsic alignment strengths in different cosmologies. We show the correlation of the two components of the halo ellipticity $\epsilon_{+}$, $\epsilon_{\times}$ with the respective tidal field components $T_{+}$, $T_{\times}$, for $N$-body, $1024^3$ resolution, $L_{\text{box}} = 40$ cMpc$/h$ runs at redshift $z=4.38$. Each dot represents one halo colour-coded by its triaxiality $T$. The blue dashed lines correspond to a linear fit to the binned data points, the shaded bands displaying the standard error on the mean (SEM) in each bin. The blue (CDM) and red solid lines (cFDM) depict the fits to the anisotropy-corrected data. All values are in units of $c^2(\mathrm{{cMpc}}/h)^2$.}
\label{f_d_fitting}
\end{figure*}

Note that the derivation of Eq. \eqref{e_lam} assumes that the objects under consideration are in a steady-state Jeans equilibrium with small perturbations. When fitting for $D$, we thus discard DM resolution elements that lie outside of the overdensity radius $R_{200}$ of the parent {\fontfamily{cmtt}\selectfont FoF} halo in an attempt to better satisfy the assumption. To reduce the bias in determining the complex ellipticity $\epsilon$, we further restrict ourselves to central subhalos, as we did when investigating mass density and shape profiles.

\subsubsection{Isotropisation of the Ellipticity Frame} 
As $\epsilon_{+}$ and $\epsilon_{\times}$ transform as components of a (spin-2) tensor, they are sensitive to the absolute orientation of structure. Since our simulation boxes of side lengths 10 and 40 cMpc/$h$ are not large enough to contain isotropic large-scale structure, cosmic variance will induce statistical differences in the parameters derived from the two components $\epsilon_{+}$ and $\epsilon_{\times}$. To circumvent this, we follow a randomisation technique outlined in \cite{Zjupa_2020}: It consists of randomising the \textit{local} $xy$-frame in which the alignment parameter $D$ is measured by rotating the frame by a random angle between $0$ and $\pi$, and averaging over the obtained results for $D$. The more randomisations one performs the more $D$ as obtained from $\epsilon_{+}$ will match the one obtained from $\epsilon_{\times}$. In practice it is sufficient to perform $10^3$ randomisations, the results of which we will denote by $D_{\text{iso}}$.

\subsubsection{Linear Alignment Model Fitting}
In Fig. \ref{f_d_fitting} we present some selected best-fit results for the alignment parameter $D$. Note that choosing a smoothing scale of $\lambda =$ 1 cMpc/$h$ while not imposing any {\fontfamily{cmtt}\selectfont FoF} halo mass cuts is slightly inconsistent, in that a smoothing scale of $\lambda =$ 1 cMpc/$h$ would suggest imposing a halo mass floor of $M_{\text{min}} \sim 10^{11} M_{\odot}/h$. In turn, $M_{\text{min}}$ determines the smoothing scale below which the matter power spectrum $P(k)$ used for computing tidal fields should be cut off, since tidal field fluctuations on scales smaller than the scale corresponding to $M_{\text{min}}$ cannot be relevant to the alignment process. The choice of the lower mass limit $M_{\text{min}}$ matters when considering averaged values of virial quantities and, consequently, of the alignment parameter, the primary reason why the employment of a mass scale of $10^{12}M_{\odot}/h$ led \cite{Tugendhat_2018} to measure a large $D$-value of $10^{-4}$ (Mpc$/h)^2$. Since our focus lies on analysing trends with the cFDM particle mass $m$ and redshift $z$, rather then exploring the whole parameter space that includes the smoothing scale $\lambda$ and halo mass $M$, our lower mass limit $M_{\text{min}}$ is merely determined by the requirement to reasonably resolve a halo. In other words, by discarding central subhalos that are comprised of fewer than $400$ particles, we obtain $M_{\text{min}} = 400 \times m_{\mathrm{DM}} = 2.0 \times 10^9 M_{\odot}/h$. For a comprehensive analysis on the subtle $\lambda$-dependence cf. \cite{Zjupa_2020}.\par 

\renewcommand{\arraystretch}{2}
\begin{table*}
	\centering
    \begin{tabular}{c | c | c c c}
    redshift & CDM & $m=2\times 10^{-21}$ eV & $m=7\times 10^{-22}$ eV & $m=10^{-22}$ eV \\ \hline \hline
    $z=10.90$ & $1.01\times 10^{-6} \pm 4\times 10^{-8}$ & $0.07 \pm 0.05$ & $0.08 \pm 0.06$ & $0.45 \pm 0.19$ \\ \hline
    $z=7.01$ & $9.59\times 10^{-7} \pm 2.4\times 10^{-8}$ & $0.06 \pm 0.04$ & $0.07 \pm 0.03$ & $0.58 \pm 0.11$ \\ \hline
    $z=4.38$ & $8.67\times 10^{-7}\pm 1.9\times 10^{-8}$ & $0.09 \pm 0.03$ & $0.12 \pm 0.04$ & $0.51 \pm 0.08$
    \end{tabular}
    \caption{Results for the isotropised alignment parameter $D_{\text{iso}}$ for various redshifts $z$ (first column), in CDM (second column) and cFDM of various particle masses $m$ (columns three, four and five), obtained by averaging the isotropised linear fits from the two components $\epsilon_{+}$, $\epsilon_{\times}$. Results are for $N$-body, $1024^3$ resolution, $L_{\text{box}} = 40$ cMpc$/h$ runs. The second column quotes $D_{\text{iso}}$ values for CDM in units of $c^2\text{(Mpc}/h)^2$, while the last three columns show the fractional differences $\left(D^{\mathrm{cFDM}}_{\text{iso}}-D^{\mathrm{CDM}}_{\text{iso}}\right)/D^{\mathrm{CDM}}_{\text{iso}}$. The redshift spacings correspond to equal logarithmic spacings in the scale factor $a$, i.e. $a_{z=7.01}/a_{z=10.90} = a_{z=4.38}/a_{z=7.01} \sim 1.49$.}
    \label{t_d_iso}
\end{table*}
\renewcommand{\arraystretch}{1}

We find that the $D$-values in Fig. \ref{f_d_fitting} that are obtained from the real part of the halo ellipticity differ from the imaginary one by at most $2.6 \sigma$. In contrast, the isotropised values $D_{\text{iso}}$ differ by at most $0.24 \sigma$. The latter ones are thus clearly more reliable, hence we present results for the isotropised values $D_{\text{iso}}$ in Table \ref{t_d_iso}, varying the redshift and the cosmology. We observe two important intrinsic alignment trends which we describe in Sections \ref{sss_z_dep}, and \ref{sss_m_dep}.\par 

\subsubsection{Increase of Alignment Strength with Redshift}
\label{sss_z_dep} 
As Table \ref{t_d_iso} suggests, we discern a strong dependence of $D_{\text{iso}}$ on cosmic time, decreasing from $D_{\text{iso}} \sim 1.01\times 10^{-6}c^2$ (cMpc/$h$)${}^2$ at redshift $z=10.90$ to about $D_{\text{iso}} \sim 8.67 \times 10^{-7}c^2$ (cMpc/$h$)${}^2$ at redshift $z=4.38$ in case of CDM. The same trend is exhibited by the cFDM cosmologies as well, with reductions in $D_{\text{iso}}$ by about $15$\% from $z=10.90$ to $z=4.38$. The origin of the $z$-dependence lies in the intricate responsivity of DM dynamics to changes in the surrounding tidal field. Such responsivity is believed to be impacted by the concentration $c$ of the halo and other halo properties, but we defer such analysis to future work. At lower redshifts, the $z$-dependence is more accentuated, with $D_{\text{iso}}$ decreasing by a factor of $\sim 3.7$ from $z=1$ to $z=0$, see \cite{Zjupa_2020} for an analogous analysis of the intrinsic alignment of the luminosity distribution of ellipticals in the $V$-band. The trend of an increasing alignment strength with redshift is also observed at low-$z$ by \cite{Samuroff_2021} in both MassiveBlack-II and the TNG300 simulation, as well as by \cite{Tenneti_2015} for the non-linear alignment model in the MassiveBlack-II simulation.\par

\subsubsection{Increase of Alignment Strength with cFDM Particle Mass}
\label{sss_m_dep} 
As we decrease the cFDM particle mass $m$, one noticeable change in Fig. \ref{f_d_fitting} comes from the higher average in the triaxiality $T$ of halos. This is in agreement with our results in Section \ref{ss_shape_stats}, where we have seen that cFDM halos are more prolate at the virial radius $R_{200}$ than their CDM counterparts. By the same token, the distributions of both the real and the imaginary part of the ellipticity $\epsilon_{\times}$ exhibit more outliers / heavier tails at larger absolute values than for CDM. This is expected, since prolate halos tend to look - on average - prolate in projection as well. This circumstance together with the distribution of local tidal field values $T_{+}$, $T_{\times}$, conspire to a strong dependence of the alignment strength on the cFDM particle mass $m$. The smaller $m$ and thus the smaller the half-mode scale $k_{1/2}$, the more $D_{\text{iso}}$ grows. The fractional difference in $D_{\text{iso}}$ between the most extreme cFDM model with $m=10^{-22}$ eV and CDM is as high as $0.45 \pm 0.19$ at $z=10.9$. At $z=4.38$, the fractional difference reads $0.51\pm 0.08$, which is different from zero at a level of about $6.4 \sigma$.\par

\section{Conclusions}
\label{s_conclusions}
In this work, we show that the impact of a small-scale cutoff in the primordial power spectrum on high-redshift virialised structures such as halos is profound, altering their density profiles, shapes and intrinsic alignment. We analyse dark matter halos in a suite of cosmological $N$-body simulations with the following specifications: two box sizes of $L_{\text{box}}=10$ and $40$ cMpc$/h$, DM resolutions of $256^3$, $512^3$ and $1024^3$, $\Lambda$CDM and classical FDM (cFDM) with particle masses $m=10^{-22}, \ 7\times 10^{-22}, \ 2\times 10^{-21}$ eV. cFDM ignores quantum pressure which is justified on the intermediate scales of $k \sim 2 - 18$ Mpc/$h$ we are interested in assuming that gravitational interactions do not re-thermalise axions. We focus on three distinct properties of central subhalos in this work:

\begin{enumerate}[i]
\item The spherically averaged internal mass distribution of dark matter halos reflects the broken hierarchy of structure formation in cFDM in that the monotonicity in the concentration-mass relation $c$-$M$ is also broken. Instead, the concentration peaks at around two orders of magnitudes above the half-mode mass $M_{1/2}$, a result that is fairly constant with redshift and in agreement with \cite{Bose_2015, Ludlow_2016}. Universality of density profiles in cFDM is at best an approximation, since concentration does not depend on halo mass close to the concentration peak in cFDM cosmologies. This is similar to the plateau regime at the high-mass end (not captured with relaxed halos) that is apparent in both CDM and cFDM cosmologies \citep{Klypin_2016}. Concentration PDFs can be well fit by lognormal distributions in CDM as well as cFDM cosmologies in the investigated redshift range of $3.4 \leq z \leq 6.3$.

\item The asymmetry-sensitive shape parameters $q$ (intermediate-to-major axis ratio) and $s$ (minor-to-major axis ratio) exhibit a monotonicity relation in $N$-body simulations of $\Lambda$CDM: Both $q$ and $s$ are monotonously increasing as a function of ellipsoidal radius for pure $N$-body simulations, a monotonicity that is thus well maintained at high redshifts in $\Lambda$CDM. It is known that baryonic physics in the centres of halos breaks the monotonicity below $z \sim 3$, with radiative feedback processes sphericalising the central regions \citep{Chua_2019, Chua_2021}. We now find that monotonicity is also broken by a small-scale cutoff in the primordial density flucutations, in models such as FDM and cFDM. In the latter, using a large sample of DM halos we show that they are significantly less spherical (lower $q$ and $s$) and more prolate (higher triaxiality $T$) than their CDM counterparts close to and beyond the virial radius $R_{\text{vir}}$. Small-mass halos experience the strongest shape distortions, mediated by the tidal fields of cosmic filaments into which many of them are embedded. Closer to the halo centre, we observe higher $q$-values for high-mass halos and in addition lower $s$-values for intermediate mass halos, translating into lower triaxialities $T$ in both cases. Likewise, the anticorrelation between sphericity and halo mass that is observed in simulations of CDM is broken in cFDM cosmologies.

\item The intrinsic alignment of halos / galaxies is both a blessing and a curse, the former due to the cosmological information that it contains and the latter by constituting a significant contaminant for past and future weak lensing surveys such as \href{http://www.euclid-ec.org/}{Euclid}. We find that geometric measures for intrinsic alignment such as the shape-position and the shape-shape alignment statistics are sensitive to the cosmological model. In particular, for small 3D halo pair separations we show that median shape-position correlations $|\cos \theta |$ increase with decreasing particle mass $m$ and in the extreme cFDM model with $m=10^{-22}$ eV climb up to $0.753^{+0.021}_{-0.010}$ as opposed to $0.653^{+0.006}_{-0.002}$ in $\Lambda$CDM, at redshift $z=4.38$ for intermediate halo masses of $10^{10}-10^{11}M_{\odot}/h$. We also carry out a linear alignment model analysis on the halos across the full mass range in the various cosmologies. For the inferred isotropised linear alignment magnitudes $D_{\text{iso}}$, we confirm the $\Lambda$CDM-trend of larger $D_{\text{iso}}$ values with increasing redshift and extend the result to cFDM. More importantly, we find very significant differences in $D_{\text{iso}}$ between the cFDM models and $\Lambda$CDM, with the $m=10^{-22}$ eV model differing from $\Lambda$CDM by as much as $6.4 \sigma$ at redshift $z=4.38$.
\end{enumerate}

We have seen that imprints of a primordial power spectrum cutoff on the internal structure and alignment of DM halos are strikingly visible upon closer look at the properties of halos in the high-$z$ Universe. In this spirit, this work attempts to improve our understanding of the high-$z$ Universe in the alternative FDM scenario. Except for the central regions of halos in which quantum pressure gives rise to de Broglie wavelength-sized solitonic cores as demonstrated in the Supplementary Materials, the conclusions in this work are expected to hold for string theory-motivated FDM cosmologies. High-redshift observations are thus a promising avenue to provide evidence for / against FDM cosmologies.\par

In particular, on scales slightly larger than those considered in this work, the next generation of line intensity mapping (LIM) surveys is expected to be instrumental in putting constraints on FDM and mixed dark matter models, especially in view of upcoming experiments such as \href{https://www.jpl.nasa.gov/missions/spherex}{SPHEREx}. While this spacecraft will observe near-infrared hydrogen recombination lines, 21 cm cosmology promises new insights at even higher redshifts well into Cosmic Dawn with the upcoming \href{https://www.skatelescope.org/the-ska-project/}{SKA} telescope. Its low-frequency component SKA1-Low will be able to rule out a $m\sim 2.6\times 10^{-21}$ eV FDM-like cosmology at more than $2 \sigma$ with 1000 hours of observations at $z \sim 5$ \citep{Carucci_2015}. Combining an SKA1-Mid-like LIM survey with the CMB measurements at the future Simons Observatory, an $m = 10^{-22}$ eV FDM model could be constrained at a few percent \citep{Bauer_2021}. Upcoming 21 cm power spectrum measurements by \href{https://reionization.org/}{HERA} will help determine the FDM particle mass to within $20$\% at $2 \sigma$ level \citep{Jones_2021}, with even better sensitivity in the mass range $10^{-25}$ eV $\leq m \leq 10^{-23}$ eV \citep{Flitter_2022}. However, it remains to show how some of these forecasts can be extended to higher redshifts and how they are modified by non-linear FDM dynamics. 

\section{Acknowledgements}
It is a pleasure to thank Shihong Liao, who has provided powerful insights into low-magnitude power spectrum estimation techniques and the setting up of initial loads for simulations with a primordial power spectrum cutoff. We further thank Debora Sijacki and Nina Sartorio for fruitful discussions. We are also grateful to Sophie Koudmani and Martin Bourne for providing help on the ins and outs of \scshape{Arepo}\normalfont . TD acknowledges support from the Isaac Newton Studentship and the Science and Technology Facilities Council under grant number ST/V50659X/1. MBK acknowledges support from NSF CAREER award AST-1752913, NSF grants AST-1910346 and AST-2108962, and HST-AR-15809, HST-GO-15658, HST-GO-15901, HST-GO-15902, HST-AR-16159, and HST-GO-16226 from the Space Telescope Science Institute, which is operated by AURA, Inc., under NASA contract NAS5-26555. AF is supported by the Royal Society University Research Fellowship. The simulations were performed under DiRAC project number ACSP253 using the Cambridge Service for Data Driven Discovery (CSD3), part of which is operated by the University of Cambridge Research Computing on behalf of the \href{https://dirac.ac.uk}{STFC DiRAC HPC Facility}. The DiRAC component of CSD3 was funded by BEIS capital funding via STFC capital grants ST/P002307/1 and ST/R002452/1 and STFC operations grant ST/R00689X/1. DiRAC is part of the National e-Infrastructure.

\section{Data Availability}
\label{s_data_availability}
We have published \href{https://github.com/tibordome/cosmic\textunderscore profiles}{\scshape{CosmicProfiles}\normalfont}, a fast, pip- and conda-installable open-source implementation of various shape and density profile estimation and fitting algorithms in Cython. Its documentation can be found \href{https://cosmic-profiles.readthedocs.io}{here}. High-level data products are available upon reasonable request.

\bibliographystyle{mnras}
\bibliography{refs}

\label{lastpage}
\end{document}


\label{firstpage}
\pagerange{\pageref{firstpage}--\pageref{lastpage}}
\maketitle

\begin{keywords} cosmology: theory, dark matter, large-scale structure of Universe
\end{keywords}



\titleformat{\section}{\normalfont\large\bfseries}{S \thesection}{1em}{}

\section{Initial Loads for Simulations with a Power Spectrum Cutoff}
\label{a_initial_loads}

For large-scale cosmological $N$-body and (magneto\=/)hydrodynamical simulations, grid and glass methods are the two most popular choices to generate uniform particle distributions. Such uniform particle distributions are then subjected to a Lagrangian perturbation theory of a certain order to generate the initial conditions (ICs) for the simulation. Popular denominations for such uniform particle distributions are \textit{pre-initial conditions} \citep[pre-ICs,][]{Baertschiger_2002, Joyce_2009} or \textit{initial loads} \citep{Jenkins_2010}. Pre-ICs should contain as little power as possible, considerably undercutting the targeted physical power spectrum on all simulated scales.\par 

\begin{figure*}
\includegraphics[scale=0.35]{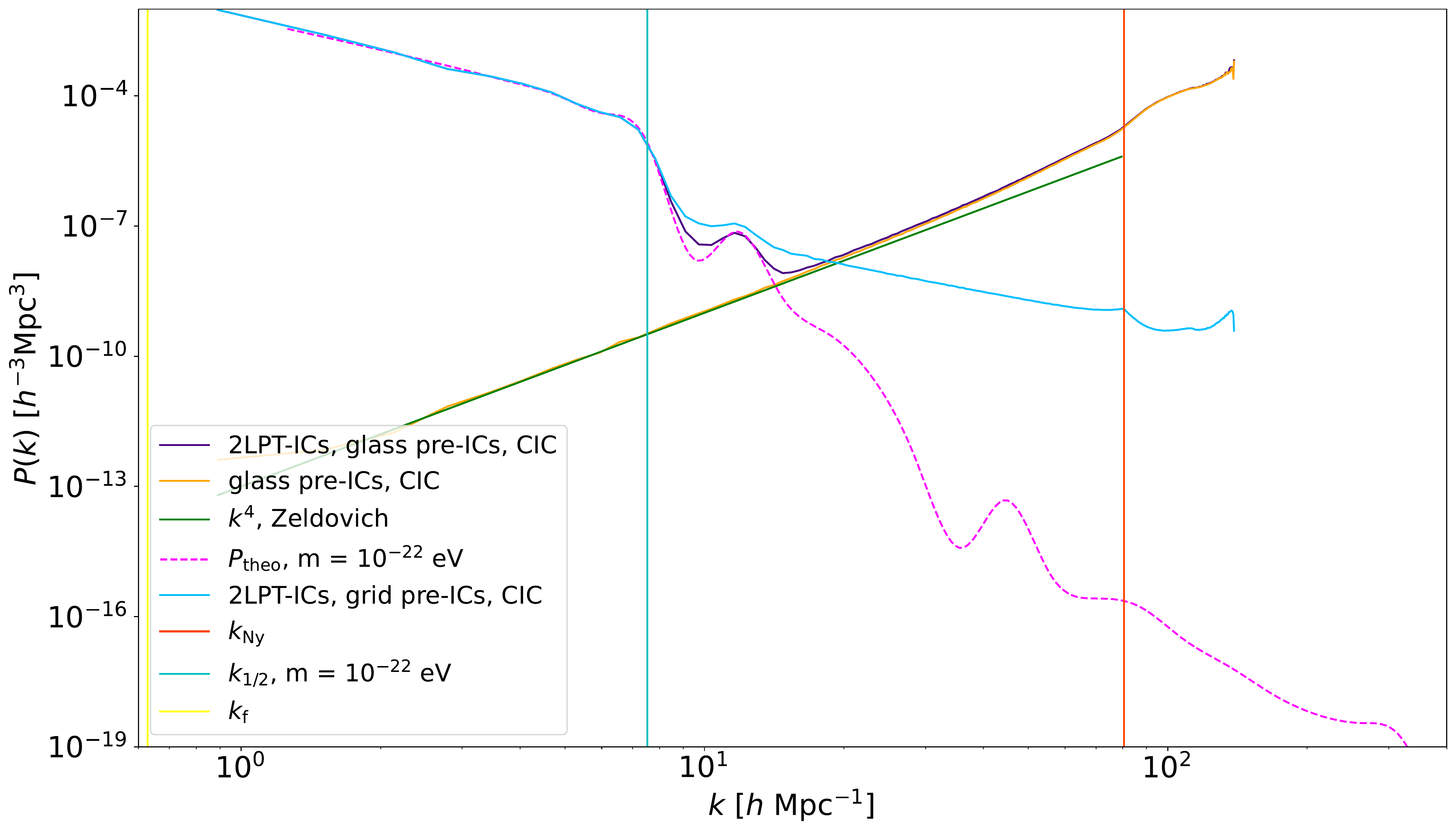}
\caption{Grid vs. glass pre-ICs power spectra. Dark purple: CIC-estimated power spectrum of the ICs that were generated with glass pre-ICs. Orange: CIC-estimated power spectrum of glass pre-ICs. Green: theoretical minimal power spectrum of $P(k) \propto k^4$. Magenta dashed: Target power spectrum for FDM with particle mass $m=10^{-22}$ eV. Blue: CIC-estimated power spectrum of the ICs that were generated with grid pre-ICs. Vertical red line: Nyquist frequency $k_{\text{Ny}}$. Vertical turquoise line: Half-mode scale $k_{1/2}$ for FDM with $m=10^{-22}$ eV. Vertical yellow line: Fundamental scale of the box $k_f$.}
\label{f_grid_vs_glass}
\end{figure*}

A uniform Poisson distribution of particle positions with its stochastic `root-N' fluctuations often exceeds the density fluctuations predicted by the desired model over a range of scales, making it unsuitable. Most early cosmological simulations in the 80s and 90s adopted a regular cubic lattice as the initial load, its symmetry ensuring no growth of structure in the absence of imposed perturbations, cf. \cite{Efstathiou_1985}. A grid is easy to produce given its computational complexity being $\mathcal{O}(N)$, $N$ being the number of particles in the grid. However, large-scale coherence and the preferred directions inherent to a grid give rise to numerical \textit{lattice effects}, which persist to low redshifts especially in low-density environments (e.g. voids) of the simulation.\par 

\cite{WhiteInSchaeffer_1996} thus suggested using glass-like pre-ICs, the dynamical equilibrium of a Poisson-random configuration of $N$ particles evolved under anti-gravity. Just as for the grid particles, the total force on each particle vanishes, with the benefit that there are no preferred directions and no long-range order. On scales much larger than the mean inter-particle spacing, the power spectrum of a well-prepared glass is close to the theoretical minimal power spectrum of $P(k) \propto k^4$, which is expected for any discrete stochastic system obeying mass and momentum conservation laws \citep{Zeldovich_1965, Peebles_1980, Baugh_1995}. Over the years, the family of pre-ICs methods has been extended by the quaquaversal tiling of space \citep{Hansen_2007} and the geometrical equilibrium obtained via constrained Voronoi tessellation (CCVT), cf. \cite{Liao_2018}.\par 

This Appendix deals with the question of optimal pre-ICs for cosmological simulations of FDM. The analysis is performed for grid and glass pre-ICs, not least because the quaquaversal tiling of space produces many more artificial structures in simulations with power spectrum cutoffs than the other methods, as found by \cite{Wang_2007}. The stability of CCVTs is comparable to that of glasses, yet by their nature of being \textit{blue noise} distributions, we find that they are both inadequate for cosmological simulations of FDM.\par 

In this comparison, we focus on initial loads containing $N = 256^3$ particles, in a box of side length $L_{\text{box}} = 1$ cMpc/$h$. Tiling is often performed to computationally simplify pre-ICs generation by tiling $N_0 < N$ particles $N_{\text{tile}}$ many times in each dimension, i.e. $N_0 \cdot N_{\text{tile}}^3 = N$. However, we find (not shown) that it can diminish the quality of glasses. While the qualitative results presented here would remain unchanged, we refrain from using tiled pre-ICs. The glass configuration is generated using \scshape{GADGET}\normalfont -2. We evolve the Poisson-random configuration under the TreePM gravity solver of \scshape{GADGET}\normalfont -2 with a comoving softening length equalling to $1/50$ of the mean inter-particle separation, as already suggested by \cite{Liao_2018}. Significantly larger or smaller softening lengths would lead to poorer glasses, by which we mean that the power spectrum would diverge stronger from the minimal $P(k) \propto k^4$, especially at high wavenumbers.\par 

The power spectrum of the glass we generated is shown in orange in Fig. \ref{f_grid_vs_glass}. It exhibits the minimal power spectrum up to $k\sim 20$ $h$/cMpc, with a bump just above the Nyquist frequency $k_{\text{Ny}} = \pi N^{\frac{1}{3}}/L_{\text{box}}$ and a spike at the sampling frequency, $2\cdot k_{\text{Ny}}$. Such power spikes are hard to avoid even with better force resolution: The large-scale particle-mesh force calculation imposes a regular `power of 2' spatial structure, reinforced by the static Barnes-Hut oct-tree which underlies the calculation of the short-range forces. The power spikes in the glass configurations reflect these structural properties of the force construction algorithm, cf. \cite{Wang_2007}.\par 

\begin{figure*}
\includegraphics[scale=0.35]{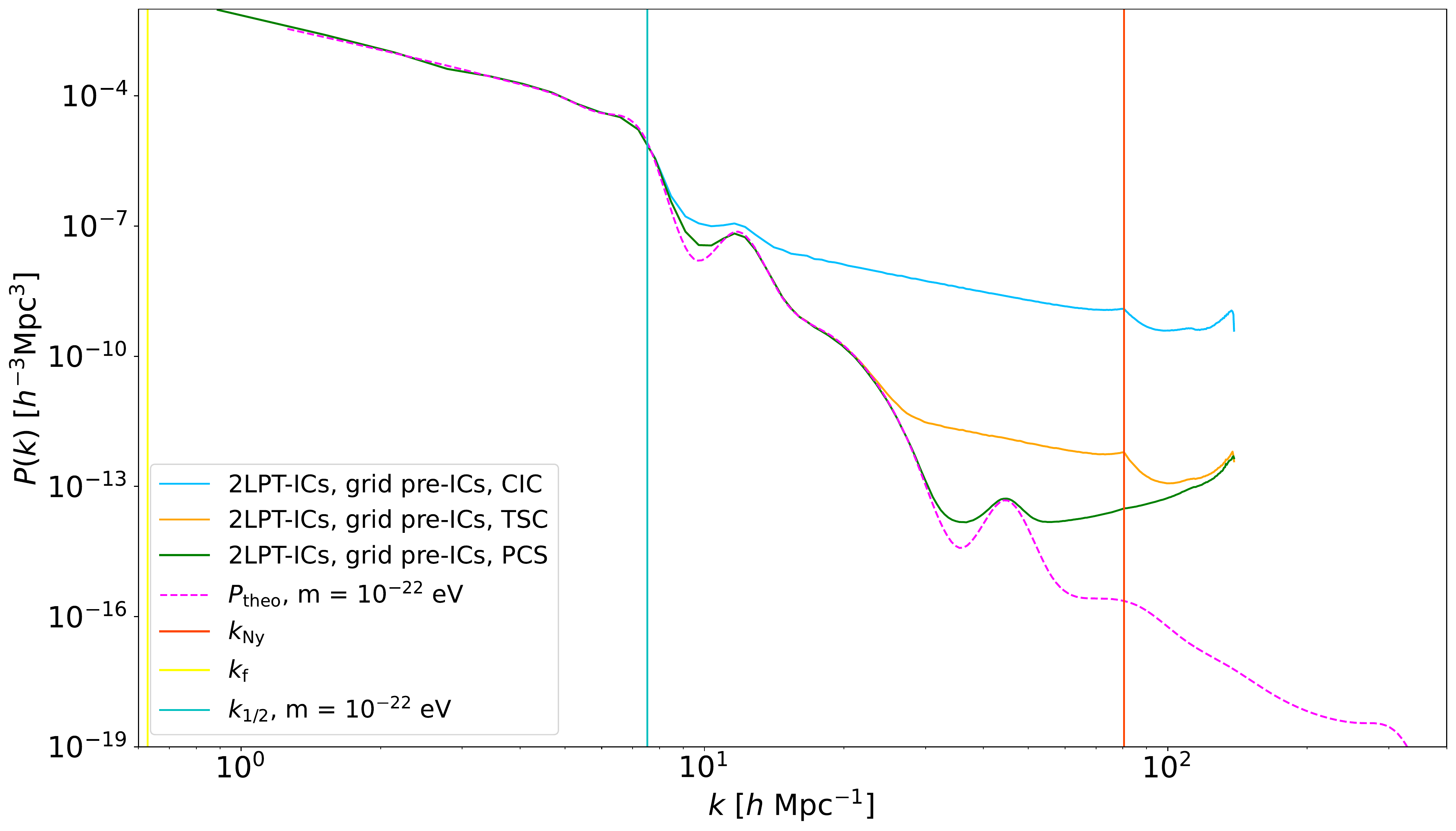}
\caption{PCS vs. CIC power spectrum estimation. Blue: CIC-estimated power spectrum of the ICs that were generated with grid pre-ICs. Orange: TSC-estimated power spectrum of the ICs that were generated with grid pre-ICs. Green: PCS-estimated power spectrum of the ICs that were generated with grid pre-ICs. Magenta dashed: Target power spectrum for FDM with particle mass $m=10^{-22}$ eV. Vertical red line: Nyquist frequency $k_{\text{Ny}}$. Vertical turquoise line: Half-mode scale $k_{1/2}$ for FDM with $m=10^{-22}$ eV. Vertical yellow line: Fundamental scale of the box $k_f$.}
\label{f_pcs_vs_cic}
\end{figure*}

\begin{figure*}
\includegraphics[scale=0.65]{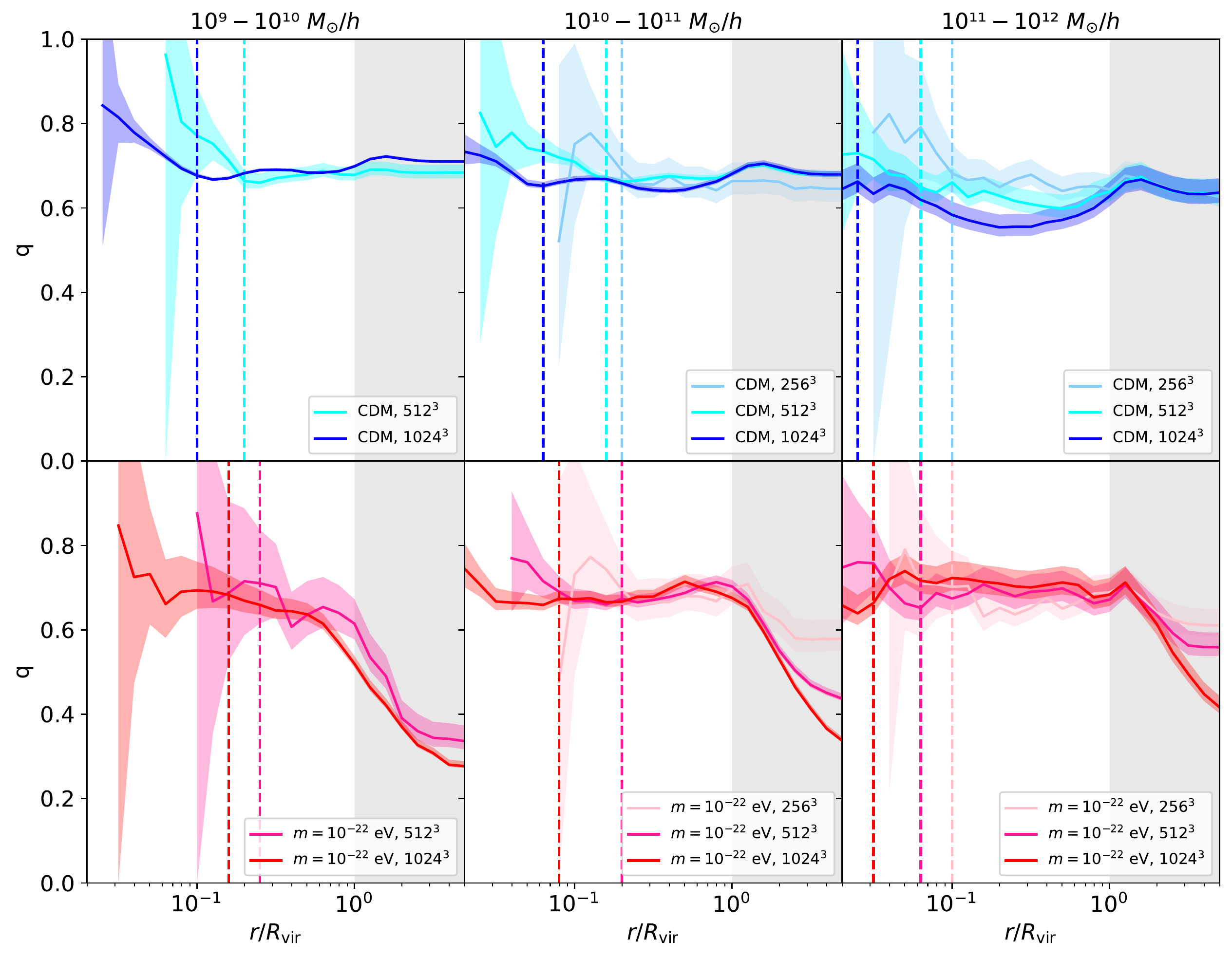}
\caption{Dependence of intermediate-to-major axis ratio $q$ profiles on resolution in different cosmologies, for $N$-body, $L_{\text{box}} = 40$ cMpc$/h$ runs at redshift $z=4.38$. Top row shows $q$-profiles of CDM halos for DM resolution of $N=256^3, 512^3, 1024^3$. Bottom: analogous profiles for halos in cFDM with particle mass $m=10^{-22}$ eV. Vertical lines demarcate $\kappa = 7$ lower convergence radii. The shaded light-grey region denotes the ambient cosmic environment of the halos.}
\label{f_conv_test}
\end{figure*}

The target power spectrum $P_{\text{theo}}$ is shown by the magenta dashed line and was generated using \scshape{AxionCamb} \normalfont \citep{Hlozek_2015} for an FDM particle mass of $m=10^{-22}$ eV. Given the target power spectrum, we use the \scshape{2lptic} \normalfont code to generate the initial conditions at $z_{\text{IC}} = 127$. We performed all tests for NGenIC \citep{Springel_2005} as well and found that the second-order Lagrangian perturbation theory in \scshape{2lptic} \normalfont \citep{Crocce_2006} gives rise to ICs power spectra that are indistinguishable from those derived from first-order Lagrangian perturbation theory in NGenIC. The ICs power spectrum obtained with glass pre-ICs is shown in dark blue in Fig. \ref{f_grid_vs_glass}, estimated using \scshape{Pylians}  \normalfont \citep{Villaescusa_2018} with a cloud-in-cell (CIC) algorithm. While the drop-off in power around the half-mode mass scale $k_{1/2}$ is well traced out, the high small-scale power of the glass is reflected in a high small-scale power of the ICs, many orders of magnitudes above the target. Consequently, glass pre-ICs and by extension all \textit{blue noise} distributions are to be avoided for cosmological simulations with power spectrum cutoffs. On the other hand, the ICs power spectrum obtained with grid pre-ICs is shown by the light-blue line. The poor carving out of the first `axion bump' around $k\sim 12$ $h$/cMpc and the spurious $-2$-like power index on small scales are both spurious features, as demonstrated in the next figure.\par

Fig. \ref{f_pcs_vs_cic} shows that with higher-order mass-assignment schemes (MASs) employed in the power spectrum estimation, better agreement is found for scales on which the magnitude of the target power spectrum is very low. While CIC constitutes a first-order MAS, the triangular shaped cloud (TSC) is second-order and the piecewise cubic spline (PCS) scheme is third-order, see \cite{Hand_2018}. We found that compensating for the MAS window function or applying interlacing techniques \citep{Hand_2018} adds little to no additional improvement. In summary, we advocate the use of grid pre-ICs in cosmological simulations featuring power spectrum cutoffs.

\section{Convergence Tests}
\label{a_conv_tests}

It is imperative to understand which regions in a given halo can be reliably resolved for the purpose of generating profiles of either density or shape. Our main guideline for assessing the inner convergence radius $r_{\text{conv}}$ relies on estimating the rescaled `relaxation' timescale $\kappa(r)$ as a function of halocentric radius,
\begin{equation}
\kappa(r) = \frac{t_{\text{relax}}(r)}{t_{\text{circ}}(R_{200})} = \frac{\sqrt{200}}{8} \frac{N(r)}{\ln N(r)}\left[\frac{\bar{\rho}(r)}{\rho_{\text{crit}}}\right]^{-1/2}.
\label{e_def_kappa}
\end{equation}
Since $\kappa(r)$ scales roughly like the enclosed number of particles $N(r)$, imposing a minimum value for $\kappa(r)$ provides a natural convergence criterion. While we opt for $\kappa(r_{\text{conv}})=1$ \citep{Power_2003} when estimating density profiles we choose $\kappa(r_{\text{conv}})=7$ for shape profiles in agreement with \cite{Vera_2011}. The more stringent requirement for halo shapes is likely a result of the 3D nature of halo shapes compared to the 1D spherically averaged mass profiles. It has been shown in \cite{Navarro_2010} that the $\kappa(r_{\text{conv}})=7$ condition also allows the circular velocity to converge to better than $2.5$\%.\par

As profiles of density are easier to converge than those of shape, we only assess shape convergence properties in the following. Furthermore, we find that it is slightly easier to make sphericity $s(r)$ converge than the intermediate to major axis ratio $q$, in agreement with \cite{Chua_2019}. Fig. \ref{f_conv_test} thus shows the dependence of $q$-shape profiles on resolution $N$, comparing $N$-body $\Lambda$CDM runs to cFDM ones with $m=10^{-22}$ eV. We vary $N$ between $N=256^3, 512^3$ and $1024^3$, and group halos into three logarithmic mass bins. The lowest-resolution runs with $N=256^3$ do not resolve any halos in the lowest mass bin of $10^9-10^{10}M_{\odot}/h$. The higher resolution runs of $N=512^3, 1024^3$ do agree well for CDM, yet show considerable discrepancy for cFDM close to the virial radius $R_{\text{vir}}$. As the small-mass cFDM halos and the cosmic filaments into which they are embedded get better resolved, $q$ gradually drops around $R_{\text{vir}}$ and is expected to saturate around $N\sim 1024^3$.\par 

Intermediate- and high-mass halos are easier to converge for both CDM and cFDM. For CDM, the trough in $q$ below the virial radius becomes more pronounced as the resolution increases, but the overall tendencies are already evident for $N=256^3$. Since the cosmic environment beyond the virial radius of halos is not virialised, the beyond-virial radius profiles are very sensitive to changes in the resolution $N$ even in the higher-mass range.\par 

At redshifts $z>4$ that we are interested in, baryons have little influence on shape profiles. Wet compaction events into self-gravitating blue nuggets in the cores of halos occur at around $z\sim 2-4$ \citep{Tomassetti_2016}. At such redshifts and lower, one indeed has to worry about underestimating stellar masses or star formation rates for low-mass halos in low-resolution runs that has been observed for \scshape{Illustris}  \normalfont \citep{Vogelsberger_2013} and various other resolution-dependent baryonic effects that are expected to co-determine halo shapes. For a thorough analysis of how galactic feedback impacts halo shapes, the reader shall be referred to \cite{Chua_2021}.\par

We remind the reader that we are using the unweighted shape tensor on enclosed ellipsoidal volumes as advocated for in \cite{Zemp_2011}. While in the latter work it was also shown that using the unweighted shape tensor on thin ellipsoidal shells constitutes an even less biased method to determine local shapes, the majority of our halos does not have sufficient resolution to warrant such a strategy.\par 

Our ellipsoidal shells have a width of 1.0 dex. We varied the width between 0.25 dex and 1.0 dex but besides increased scatter did not find appreciable effects on the median shape profiles and their associated convergence. Our shape profiles are implicitly assuming a removal of substructure. We remove all satellites in a given {\fontfamily{cmtt}\selectfont FoF} halo and only retain the most massive subhalo, also called central. It has been found that substructure has a noticeable effect only near the virial radius, decreasing sphericity and increasing the prolateness of halos \citep{Chua_2019}, yet we do not pursue this avenue here.

\section{Quantum Pressure Induced Halo Structure Modifications}
\label{a_quantum_pressure}

A cutoff in the transfer function is elongating structure and modifying shapes irrespective of quantum pressure considerations, and this Appendix is devoted to illustrating this claim.\par 

To this end, we compare the high-resolution CDM, cFDM and FDM simulation triplet that constitutes the basis of the \cite{Mocz_2019} and \cite{Mocz_2020} investigations. The simulation box size is $L_{\text{box}}=1.7$ cMpc/$h$ and the FDM particle mass $m=2.5 \times 10^{-22}$ eV. Resolution varies between $N=1024^3$ for FDM and $N=512^3$ for CDM and cFDM. This combination allows to resolve superfluid velocities up to $\varv/\sqrt{a}\sim 250$ km/s, where $a$ is the scale factor and $\varv$ the peculiar velocity, a soliton core for a halo of mass $M_h \sim 10^8 M_{\odot}/h$, all down to redshift $z\sim 6$. Recall that in FDM, the velocity field is a gradient flow, i.e. without resolving velocities, we might get spurious structures due to quantum effects.\par 

\begin{figure}
\hspace{-0.4 cm}
\includegraphics[scale=0.55]{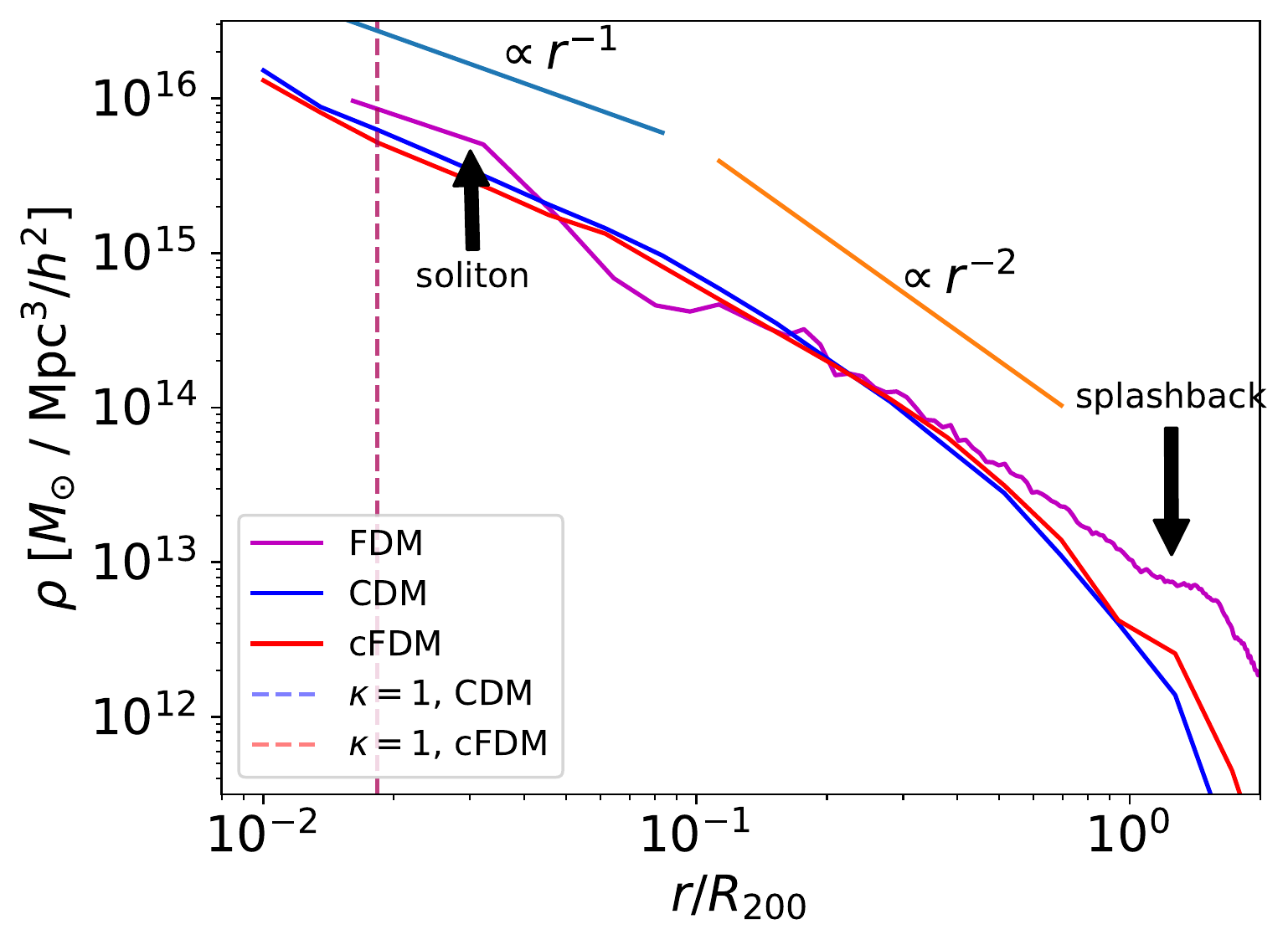}
\caption{Density profiles in CDM (blue), FDM (purple) and cFDM (red) with particle mass $m=2.5 \times 10^{-22}$ eV for the largest halo of mass $\sim 10^{10}M_{\odot}/h$ at $z\sim 6$. Dashed blue and red lines delineate $\kappa = 1$ lower convergence radii for CDM and cFDM, respectively. The $r^{-1}$ and $r^{-2}$ power laws trace two characteristic regimes of an NFW-like profile.}
\label{f_rho_prof_cdm_fdm_cfdm}
\end{figure}

The density profiles for the largest halo in the box at $z\sim 6$ are shown in Fig. \ref{f_rho_prof_cdm_fdm_cfdm}. The CDM and cFDM profiles were obtained by brute-force radial binning around the mode of the halo, as in Section 4.2, while the FDM profile was obtained by linearly interpolating the spherically averaged density to all halocentric radii of interest. Note that the modified density profile estimation is unavoidable given that the spectral solver that evolves FDM each step is solved on a grid, cf. \cite{Mocz_2017}. We observe a characteristic Einasto-like profile for all three DM prescriptions, while the NFW-motivated power-laws $r^{-1}$  at small radii and $r^{-2}$ at intermediate radii also fare well in approximating the inferred profiles. As we find in Section 4.2, cFDM halos typically feature lower concentrations than CDM halos. The red cFDM curve in Fig. \ref{f_rho_prof_cdm_fdm_cfdm} indeed emerges above the blue CDM one around $r/R_{200}\sim 0.2$, indicating a lower concentration.\par 

\begin{figure}
\includegraphics[scale=0.55]{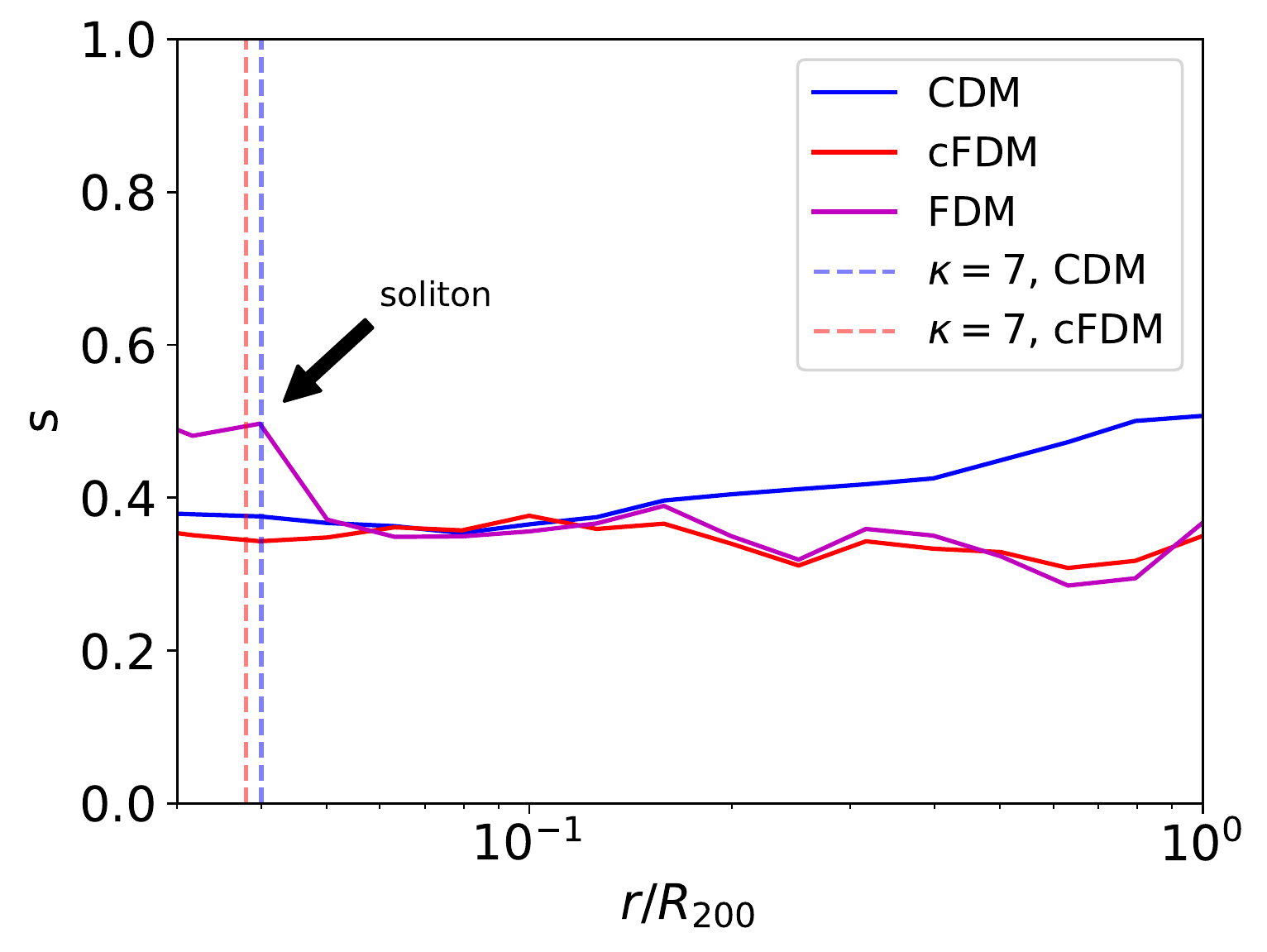}
\caption{Minor-to-major axis shape profiles in CDM (blue), FDM (purple) and cFDM (red) with particle mass $m=2.5 \times 10^{-22}$ eV for the largest halo of mass $\sim 10^{10}M_{\odot}/h$ at $z\sim 6$. Dashed blue and red lines delineate $\kappa = 7$ lower convergence radii for CDM and cFDM, respectively. The $r^{-1}$ and $r^{-2}$ power laws trace two characteristic regimes of an NFW-like profile.}
\label{f_s_prof_cdm_fdm_cfdm}
\end{figure}

On top of the density profile modifications that stem from a power spectrum cutoff, the purple FDM profile features a marginally resolved soliton core below $r/R_{200}\sim 0.03$. While this feature alters the innermost density profile, the result that cFDM halos are less concentrated than CDM ones is expected to be transferrable to FDM too. The increased resolution of the simulation triplet compared to the large-box ones in the rest of this work allows to visualise the splashback radius, the radius where particles reach the apocentre of their first orbit. At the same time, it represents a density caustic \citep{More_2015}. For this halo, the splashback radius is about 1.5 times larger then $R_{200} \sim 50$ kpc/$h$, though in general depends to lowest order on the mass accretion rate. It is more pronounced for the FDM halo possibly due to higher resolution ($N=1024^3$ vs $N=512^3$).\par 

For the very same halo of mass $\sim 10^{10}M_{\odot}/h$ at $z\sim 6$, we present shape profiles in Fig. \ref{f_s_prof_cdm_fdm_cfdm}. While all three shape profiles are obtained using the Katz-Dubinski iteration method applied on the unweighted shape tensor as described in Section 3.1, the FDM density grid first needs to be interpolated accordingly. In accordance with Section 4.3, we again find a monotonically increasing CDM minor-to-major axis profile, attaining the value of $s\sim 0.51$ at $r/R_{200}=1.0$. Both cFDM and FDM exhibit a value of $s\sim 0.35$ at the virial radius. These values agree well with those found in \cite{Mocz_2020}, but this work extends the shape estimates to lower ellipsoidal radii $r/R_{200}$. The cFDM and FDM halo remain less spherical than the CDM one down to $r/R_{200}=0.15$. Differences between cFDM and FDM become apparent below $r/R_{200}=0.05$ only, where the FDM halo features a marginally resolved soliton, as already indicated in Fig. \ref{f_rho_prof_cdm_fdm_cfdm}.

\bibliographystyle{mnras}
\bibliography{refs}

\label{lastpage}